\documentclass{aa}                                              

\usepackage[english]{babel}                                     
\usepackage[varg]{txfonts}                                      
\usepackage{upgreek}                                            

\usepackage{balance}                                            
\usepackage{multicol}                                           
\usepackage{changes}                                            
\usepackage{lscape}                                             
\usepackage{enumitem}                                           

\usepackage{orcidlink}                                          

\usepackage{graphicx}                                           
\usepackage{float}                                              
\usepackage{placeins}                                           
\usepackage{caption}                                            
\graphicspath{{Figures/}}                                       

\usepackage{siunitx}                                            
\usepackage{amsmath,amsfonts,amssymb}                           
\usepackage{xparse}                                             
\usepackage[version=4]{mhchem}                                  

\usepackage{subcaption}                                         
\usepackage{multirow}                                           
\usepackage{tablefootnote}                                      

\usepackage{natbib}                                             
\usepackage{aas_macros}                                         
\bibpunct{(}{)}{;}{a}{}{,}                                      

\usepackage{hyperref}                                           
\hypersetup{
    colorlinks=true, 
    linkcolor=blue, 
    urlcolor=blue, 
    citecolor=blue
}

\begin{document}

\newcommand{\kmps}{\si{\kilo\metre\per\second}}                     
\newcommand{\mps}{\si{\metre\per\second}}                           
\newcommand{\rads}{\si{\rad\per\second}}                            
\renewcommand{\K}{\si{\kelvin}}                                     
\newcommand{\dex}{\mathrm{dex}}                                     
\newcommand{\cte}{\mathrm{cte}}                                     

\newcommand{\Msol}{\mathrm{M_\odot}}                                
\newcommand{\Rsol}{\mathrm{R_\odot}}                                
\newcommand{\Zsol}{\mathrm{Z_\odot}}                                
\newcommand{\Lsol}{\mathrm{L_\odot}}                                

\renewcommand{\deg}{^\circ}                                         
\renewcommand{\arcmin}{^\prime}                                     
\renewcommand{\arcsec}{^{\prime\prime}}                             

\newcommand{\Teff}{T_\mathrm{eff}}                                  
\newcommand{\Teffsph}{T_\mathrm{eff,\,0}}                           
\newcommand{\Teffrot}{T_\mathrm{eff,\,rot}}                         
\newcommand{\Teffpole}{T_\mathrm{eff,\,p}}                          
\newcommand{\Teffeq}{T_\mathrm{eff,\,eq}}                           
\newcommand{\Teffavg}{T_\mathrm{eff,\,avg}}                         

\newcommand{\geff}{\text{\textit{g}}_{\rm eff}}                     
\newcommand{\geffpole}{\text{\textit{g}}_{\rm eff,\,p}}             
\newcommand{\geffeq}{\text{\textit{g}}_{\rm eff,\,eq}}              
\newcommand{\geffavg}{\text{\textit{g}}_{\rm eff,\,avg}}            
\newcommand{\logg}{\log{\text{\textit{g}}}}                         
\newcommand{\loggeff}{\log{\text{\textit{g}}_{\rm eff}}}            
\newcommand{\loggeffpole}{\log{\text{\textit{g}}_{\rm eff,\,p}}}    
\newcommand{\loggeffeq}{\log{\text{\textit{g}}_{\rm eff,\,eq}}}     
\newcommand{\loggeffavg}{\log{\text{\textit{g}}_{\rm eff,\,avg}}}   

\newcommand{\Ilambmu}{I\left(\lambda,\mu\right)}                    
\newcommand{\Frad}{F_\mathrm{rad}}                                  
\newcommand{\Lrot}{L_\mathrm{rot}}                                  
\newcommand{\Lrotobs}{L_\mathrm{rot,\,obs}}                         
\newcommand{\Eddfactore}{\Gamma_\mathrm{e}}                         

\newcommand{\MH}{\mathrm{\left[M/H\right]}}                         
    
\newcommand{\Mstar}{M_*}                                            
\newcommand{\Rpole}{R_\mathrm{p}}                                   
\newcommand{\Req}{R_\mathrm{eq}}                                    
\newcommand{\Reqcrit}{R_\mathrm{\,eq}^\mathrm{\,crit}}              
\newcommand{\Surfrot}{A_\mathrm{rot}}                               
\newcommand{\Rsurf}{R_\mathrm{eqv-a}}                               
\newcommand{\Volrot}{V_\mathrm{rot}}                                
\newcommand{\Reqv}{R_\mathrm{eqv-v}}                                

\newcommand{\vsini}{v\sin{i}}                                       
\providecommand{\vsin}[1]{$v\sin{i} = #1\,\kmps$}                   
\newcommand{\vrot}{v_\mathrm{rot}}                                  
\newcommand{\angularrot}{\Omega_\mathrm{rot}}                       
\newcommand{\periodrot}{P_\mathrm{rot}}                             
\newcommand{\vrotcrit}{v_\mathrm{crit}}                             
\newcommand{\angularrotcrit}{\Omega_\mathrm{crit}}                  
\newcommand{\periodrotcrit}{P_\mathrm{crit}}                        
\newcommand{\vrotrate}{\omega}                                      
\newcommand{\angularrotrate}{\omega_\Omega}                         
\newcommand{\periodrotrate}{\omega_P}                               
\newcommand{\inclination}{i}                                        

\NewDocumentCommand{\spectype}{m m o}                               
{#1#2\IfValueT{#3}{\,#3}}                                           

\NewDocumentCommand{\specline}{m m o}                               
    {$\mathrm{#1}\,\textsc{#2}$\IfValueT{#3}{\,$\lambda#3$}}        
\NewDocumentCommand{\multispecline}{m m m}                          
    {$\mathrm{#1}\,\textsc{#2}\,\lambda\lambda#3$}                  
\NewDocumentCommand{\ratio}{m m o m m o}                            
    {\specline{#1}{#2}[#3]$\,/\,$\specline{#4}{#5}[#6]}             

\newcommand{\FASTWIND}{\textsc{fastwind}}                           
\newcommand{\TLUSTY}{\textsc{tlusty}}                               
\newcommand{\Kurucz}{\textsc{kurucz}}                               
\newcommand{\SPAMMS}{\textsc{spamms}}                               
\newcommand{\PRISMAS}{\textsc{prismas}}                             
\newcommand{\PANORAMA}{\textsc{panorama}}                           
\newcommand{\synple}{\textsc{synple}}                               
\newcommand{\SYNSPEC}{\textsc{synspec}}                             
\newcommand{\PHOEBE}{\textsc{phoebe}}                               
\newcommand{\Python}{Python}                                        
\newcommand{\IDL}{IDL}                                              

\providecommand{\abs}[1]{\lvert#1\rvert}                            
\providecommand{\derivative}[3][]                                   
    {\frac{\mathrm{d}^{#1}\ \!#2}{\mathrm{d}\ \! #3^{#1}}}


\title
    {
    \SPAMMS: 3D spectroscopic modeling of stellar surfaces
    }

\subtitle
    {
    II. Implementation of Kurucz and TLUSTY model atmospheres
    }

\author
    {
    D. Gal{\'a}n-Di{\'e}guez \inst{1, 2}\fnmsep\thanks{Corresponding author: \texttt{dgalandieguez.astro@gmail.com}}\,\orcidlink{0000-0001-6191-8251} \and
    M. Abdul-Masih \inst{1,2}\,\orcidlink{0000-0001-6566-7568} \and
    C. Allende Prieto \inst{1,2}\,\orcidlink{0000-0002-0084-572X} \and \\
    S.R. Berlanas \inst{1,2}\,\orcidlink{0000-0002-2613-8564} \and
    A. Herrero \inst{1,2}\,\orcidlink{0000-0001-8768-2179} \and
    H. Sana \inst{3,4}\,\orcidlink{0000-0001-6656-4130}
    }

\institute
    {
    Instituto de Astrofísica de Canarias, c/ Vía Láctea, s/n, E-38205 La Laguna, Tenerife, Spain \and
    Departamento de Astrofísica, Universidad de La Laguna, E-38206 La Laguna, Tenerife, Spain \and
    Institute of Astronomy, KU Leuven, Celestijnlaan 200D, 3001 Leuven, Belgium \and
    Leuven Gravity Institute, KU Leuven, Celestijnenlaan 200D, box 2415, 3001 Leuven, Belgium
    }

\date
    {
    Received month day year; accepted month day year
    }

\abstract
    {
    Accurate modeling of stellar spectra is essential for deriving the physical properties of stars. However, traditional model atmospheres often oversimplify key phenomena -- such as deformations due to rotation or multiplicity -- that break spherical symmetry. The Spectroscopic PAtch Model for Massive Stars (\SPAMMS) addresses these limitations by explicitly accounting for surface distortions, resulting in more realistic spectra for deformed stars. Yet, \SPAMMS{} has been limited by the spectral types covered by its model atmosphere grids.
    }
%
%
    {
    We aim to extend the parameter space of the stellar atmosphere grids available to the \SPAMMS{} framework. By expanding the coverage of effective temperatures, surface gravities, and metallicities, we enable the synthesis of spectra across a broader range of spectral types.
    }
%
%
    {
    We computed specific intensities, $\Ilambmu$, for $101$ emergent angles using \PRISMAS{} (Pipeline of Radiative Intensity Synthesis for Meshed Atmospheric Surfaces). Built upon \synple{}, \PRISMAS{} employs precomputed local thermodynamic equilibrium (LTE) and non-LTE model atmospheres, including two ATLAS9-Kurucz grids and the TLUSTY-based OSTAR2002 and BSTAR2006 models.
    }
%
%
    {
    The computed intensity grids span effective temperatures from $3500$ to $55000\,\K$ and surface gravities between $0.0$ and $5.0\,\dex$ -- covering the entire range of spectral types from O- to K-type stars. These grids also incorporate multiple metallicities ($0-30\,\Zsol$) and microturbulent velocities ($1$,~$3$,~$5$,~$10\,\kmps$), enabling detailed modeling of stars across diverse evolutionary stages. The spectral coverage extends from $3000$ to $9000\,\AA$, spanning both the ultraviolet and optical regimes with a sampling of $\Delta \lambda = 0.01\,\AA$. As a proof of concept, we adopted the new intensity grids to model (i) a rapidly rotating B-type star, and (ii) an eclipsing Algol-type binary using \SPAMMS{}.
    }
%
%
    {
    The implementation of LTE-Kurucz and NLTE-TLUSTY model atmospheres significantly expands the parameter space accessible to \SPAMMS{}. The code can now generate realistic synthetic spectra for a wide range of stellar types and geometries, including nonspherical and multiple systems. Therefore, by explicitly modeling three-dimensional surface effects, \SPAMMS{} reproduces key observational signatures across a broad range of stellar masses and effective temperatures -- from O- to K-type stars.
    }
%
%
\keywords
    {
    stars: early-type -- stars: late-type -- stars: atmospheres --  stars: rotation -- binaries: general -- techniques: spectroscopic    
    }

\maketitle
\nolinenumbers

\section{Introduction} \label{S: intro}
    {
    Accurate computation of stellar parameters depends critically on detailed model atmospheres. Such models describe the interaction between radiation and matter in the outer layers of stars, and provide the corresponding temperature, density, and pressure stratification \citep{2008oasp.book.....G}.

    However, modeling stellar atmospheres remains a complex task. It requires an accurate treatment of key quantities essential for radiative transfer, which are highly sensitive to atomic and molecular data (e.g., occupation numbers and opacities). An additional complication stems from the assumption of local thermodynamic equilibrium (LTE). While computationally efficient, this approximation breaks down in the atmospheres of hot, massive stars. In these regimes, non-LTE effects dominate the atomic populations and fundamentally shape the emergent radiation field (see \citealt{2015tsaa.book.....H} and \citealt{2019A&A...621A..85H} for a review and references therein).

    Beyond these intrinsic challenges, geometric and dynamical effects further complicate stellar modeling. In particular, rotation and multiplicity strongly influence stellar structure and emergent spectra \citep{2024arXiv240903329P}. Rapid rotation causes centrifugal flattening, producing significant temperature and brightness variations across the surface \citep{1924MNRAS..84..665V, 2011A&A...533A..43E}. Similarly, stellar multiplicity distorts the combined spectrum through mutual irradiation, tidal deformation, and line blending \citep{2013A&A...552A..39P, 2024arXiv241016114S}. Despite their importance, current stellar models often neglect or oversimplify these effects due to computational constraints.
    
    Furthermore, to account for these phenomena, traditional modeling approaches adopt several simplifications. Rotational effects are commonly mimicked by convolving a one-dimensional (1D), ``nonrotating'' synthetic spectrum with a rotational kernel \citep{1933MNRAS..93..478C, 2008oasp.book.....G}. However, this first-order approximation fails to capture inherently 3D effects, such as surface flux anisotropies \citep{1924MNRAS..84..665V, 2023A&A...669L..11A}. 
    
    In multiple systems, stellar components are usually modeled as independent sources \citep[e.g.,][]{2019ApJ...880..115A}. Additionally, disentangling techniques are applied to isolate the spectral contribution of each star \citep{2009A&A...494..399H, 2024MNRAS.530.1935S, 2026arXiv260402111M}. Yet, these methods assume that the spectra of the components can be independently analyzed, neglecting radiative and dynamical interactions within the system --~which can significantly modify the observed composite spectrum \citep{2021A&A...651A..96A}.

    In contrast, \SPAMMS{} \citep[Spectroscopic PAtch Model for Massive Stars,][]{2020A&A...636A..59A} is a spectral synthesis code designed to model distorted stellar surfaces in 3D. First, it uses \PHOEBE{} \citep[PHysics Of Eclipsing BinariEs,][]{2016ApJS..227...29P} to represent the stellar geometry as a mesh of discrete surface elements, each with a local effective temperature ($\Teff$) and surface gravity ($\logg$). Then, \SPAMMS{} assigns to each patch of the mesh emergent specific intensities, which are integrated over all visible surface regions. This procedure consistently accounts for the complex surface morphology, incorporating gravity darkening, temperature gradients, surface velocity fields, and stellar deformation. Consequently, \SPAMMS{} captures the effects of rotation and tidal distortions on the emergent spectra of rapidly rotating stars and interacting binaries.

    However, the current implementation of \SPAMMS{} is limited by the available stellar model atmospheres. Earlier versions relied exclusively on \FASTWIND{} \citep[Fast Analysis of STellar atmospheres with WINDs,][]{1997A&A...323..488S, 2005A&A...435..669P, 2011A&A...536A..58R, 2020A&A...642A.172P}, which restricted \SPAMMS{} to O- and early B-type stars. Thus, the code could not be applied consistently to cooler stars or broader wavelength domains. 
    
    To address these limitations, we generated new intensity grids for \SPAMMS{} using precomputed atmospheres. We used both LTE and non-LTE plane-parallel models, including two ATLAS9-Kurucz grids \citep{2003IAUS..210P.A20C, 2012AJ....144..120M} and the TLUSTY-based OSTAR2002 \citep{2003ApJS..146..417L} and BSTAR2006 \citep{2007ApJS..169...83L} datasets. Hence, our new grids span a wide range of effective temperatures ($3500-55000\,\K$) and surface gravities ($0.0-5.0\,\dex$), covering spectral types from O to K. Our models also include multiple metallicities ($0-30\,\Zsol$) and microturbulent velocities ($1$,~$3$,~$5$,~$10\,\kmps$), enabling detailed modeling of an extensive variety of stellar atmospheres.

    This paper is organized into seven sections. In Sect.~\ref{S: stellar_models}, we describe the precomputed model atmospheres adopted in this work. Section~\ref{S: PRISMAS} details the computation of angle-dependent emergent intensities using the \PRISMAS{} pipeline. In Sect.~\ref{S: validation} we validate the internal consistency of the flux integration in \SPAMMS{}. Section~\ref{S: applications} demonstrates the capabilities of \SPAMMS{} by modeling the spectra of a rapidly rotating B-type star and an eclipsing Algol-type binary. Finally, Sect.~\ref{S: conclusions} summarizes the main results of this work.
    }

\section{Stellar models} \label{S: stellar_models}
    {
    We relied on precomputed model atmospheres to generate the intensity grids for \SPAMMS{}. In fact, these models provide the atmospheric structure and chemical composition required  to solve the radiative transfer equation (RTE) and compute specific intensities.
    
    To cover the full stellar parameter space -- from O- to K-type stars -- we employed both LTE and non-LTE grids. In particular, we adopted the ATLAS9-Kurucz models (Sect.~\ref{S: ATLAS9}) and the TLUSTY-based OSTAR2002 and BSTAR2006 datasets (Sect.~\ref{S: TLUSTY}).

    \subsection{LTE model atmospheres: The ATLAS9 grids} \label{S: ATLAS9}
        {
        \subsubsection{The Castelli\&Kurucz grid} \label{S: ATLAS9_Kurucz}
            {
            We adopted the ATLAS9 grid from \citet{2003IAUS..210P.A20C}, computed with the 1D Kurucz code \citep{1979ApJS...40....1K}. These models assume LTE and hydrostatic equilibrium, using a uniform set of plane-parallel layers. Convection is included via a mixing-length parameter of $\ell/H_\mathrm{p} = 1.25$\footnote{$H_\mathrm{p}$ denotes the local pressure scale height.}, while overshooting is disabled.
            
            This ATLAS9 grid improves previous ATLAS versions in two key aspects. First, models were computed with updated solar abundances from \citet{1998SSRv...85..161G}. Second, the grid included revised line opacity calculations using opacity distribution functions \citep[ODF;][]{2005MSAIS...8...34C, 2005MSAIS...8...14K}. Consequently, these changes yielded more accurate temperature and pressure stratifications and improved the line blanketing treatment.
                 
            Regarding the parameter space, the Castelli\&Kurucz grid includes effective temperatures ranging from $\Teff = 3500\,\K$ to $50000\,\K$ (Table~\ref{T: grids_params}). Temperature sampling is divided into two regimes: models are computed in steps of $250\,\K$ for $3500 \le \Teff\,\left[\K\right] \le 13000$, while increasing the spacing to $1000\,\K$ for $\Teff > 13000\,\K$. The corresponding surface gravities range from $\logg = 0.0$ to $5.0$, with a uniform sampling interval of $0.5\,\dex$. The minimum $\logg$ value depends on temperature, and is set by the lowest value that ensures gravitational stability. The grid further covers metallicities between $\MH = -2.5$ and $+0.5$ in logarithmic steps of $0.5\,\dex$. It also includes enhanced models, in which the $\alpha$ elements (O, Ne, Mg, Si, S, Ar, Ca, and Ti) are increased by $+0.4\,\dex$ relative to the solar abundance scale.
            }

        \subsubsection{The M{\'e}sz{\'a}ros grid} \label{S: ATLAS9_Meszaros}
            {
            \begin{table*}[]
                \caption{
                    Parameter space covered by the precomputed model atmospheres used in this work.
                } 
                \label{T: grids_params}
                \centering   
                \small
                \begin{tabular}{cccccccccc}
                    \hline
                    \hline \\ [-1.8ex]
                    \multicolumn{2}{c}{\multirow{2}{*}{Stellar grid}}     & \multicolumn{2}{c}{$\Teff$ $\left[\K\right]$} & \multicolumn{2}{c}{$\logg$}           & \multicolumn{4}{c}{Chemical abundances}                                                                                                                                     \\
                    \multicolumn{2}{c}{}                                  & Range                     & Step              & Range       & Step                    & $\MH$                                      & $\mathrm{\left[\alpha/H\right]}$           & $\mathrm{\left[C/H\right]}$                & $Z/\Zsol$                            \\ \hline \\ [-1.8ex]
                    \multicolumn{2}{c}{\multirow{2}{*}{Castelli\&Kurucz}} & $3500-13000$              & $250$             & $0.0-5.0$   & \multirow{2}{*}{$0.5$}  & \multirow{2}{*}{$\left(-2.5, +0.5\right)$} & \multirow{2}{*}{$+0.0$, $+0.4$}            & \multirow{2}{*}{$-$}                       & \multirow{2}{*}{$-$}                 \\
                    \multicolumn{2}{c}{}                                  & $13000-50000$             & $1000$            & $2.0-5.0$   &                         &                                            &                                            &                                            &                                      \\ \hline \\ [-1.8ex]
                    \multicolumn{2}{c}{\multirow{3}{*}{M{\'e}sz{\'a}ros}} & $3500-12000$              & $250$             & $0.0-5.0$   & \multirow{3}{*}{$0.5$}  & \multirow{3}{*}{$\left(-5.0,+1.5\right)$}  & \multirow{3}{*}{$\left(-1.5, +1.0\right)$} & \multirow{3}{*}{$\left(-1.5, +1.0\right)$} & \multirow{3}{*}{$-$}                 \\
                    \multicolumn{2}{c}{}                                  & $12000-20000$             & $500$             & $3.0-5.0$   &                         &                                            &                                            &                                            &                                      \\
                    \multicolumn{2}{c}{}                                  & $20000-30000$             & $1000$            & $4.0-5.0$   &                         &                                            &                                            &                                            &                                      \\ \hline \\ [-1.8ex]
                    \multirow{2}{*}{TLUSTY}          & OSTAR2002          & $27500-55000$             & $2500$            & $3.00-4.75$ & \multirow{2}{*}{$0.25$} & \multirow{2}{*}{$-$}                       & \multirow{2}{*}{$-$}                       & \multirow{2}{*}{$-$}                       & \multirow{2}{*}{$\left(0, 2\right)$} \\
                                                     & BSTAR2006          & $15000-30000$             & $1000$            & $1.50-4.75$ &                         &                                            &                                            &                                            &                                      \\ \hline
                \end{tabular}
                \tablefoot{
                    From top to bottom: the ATLAS9-Castelli\&Kurucz \citep{2003IAUS..210P.A20C}, ATLAS9-M{\'e}sz{\'a}ros \citep{2012AJ....144..120M}, and TLUSTY-based \citep{2003ApJS..146..417L, 2007ApJS..169...83L} grids. Columns list the effective temperature ($T_{\mathrm{eff}}$), surface gravity ($\logg$), and chemical abundances. For the ATLAS9 grids, specific enhancements in carbon ($\mathrm{\left[C/H\right]}$) and $\alpha$-elements ($\mathrm{\left[\alpha/H\right]}$) are indicated. Ranges are expressed as $\left(\text{min, max}\right)$ or with a $-$ symbol.
                    }
            \end{table*}
            
            The model atmospheres of \citet{2012AJ....144..120M} are based on the same ATLAS9 framework as the \citet{2003IAUS..210P.A20C} grid: the computations adopt a 1D, plane-parallel geometry, assuming LTE and hydrostatic equilibrium. Convection is also parameterized with a mixing-length parameter of $\ell/H_\mathrm{p} = 1.25$, while convective overshooting is switched off.

            However, the M{\'e}sz{\'a}ros grid introduces several updates relative to previous ATLAS9 releases. The new models include a corrected \ce{H2O} molecular line list, which improves the treatment of molecular opacities at low $\Teff$. The grid also provides an extended range of carbon abundances and $\alpha$-element enhancements and adopts the updated solar abundances of \citet{2009ARA&A..47..481A}. 

            But the new ATLAS9 models cover a more restricted range of effective temperatures than the Castelli\&Kurucz implementation (Fig.~\ref{Fig: grids_parameters}, middle panel). Models span from $\Teff = 3500\,\K$ to $30000\,\K$, with different temperature sampling: (i) the grid is computed with a step of $\Delta \Teff = 1000\,\K$ for $\Teff \geq 20000\,\K$; (ii) the spacing decreases to $500\,\K$ for the interval $12000 < \Teff\,\left[\K\right] < 20000$; and (iii) a finer resolution of $\Delta \Teff =250\,\K$ is adopted for $\Teff \leq 12000\,\K$. Yet, the surface gravities also range from $\logg = 0.0$ to $5.0$, with uniform steps of $0.5\,\dex$. Regarding abundances, the chemical parameter space extends from $\MH = -5.0$ to $+1.5$, including variations in $\mathrm{\left[C/H\right]}$ and $\mathrm{\left[\alpha/H\right]}$ between $-1.5$ and $+1.0\,\dex$. As a result, the M{\'e}sz{\'a}ros grid significantly extends the coverage of the metal-poor regime compared to the Castelli\&Kurucz models.
            }
        }

    \subsection{Non-LTE model atmospheres: The TLUSTY OSTAR2002 and BSTAR2006 grids} \label{S: TLUSTY}
        {
        TLUSTY computes 1D, plane-parallel stellar atmospheres in hydrostatic and radiative-convective equilibrium \citep{2017arXiv170601859H, 2017arXiv170601935H, 2017arXiv170601937H, 2021arXiv210402829H}. Moreover, it solves the equations of radiative transfer and statistical equilibrium under non-LTE conditions. This approach is crucial for hot stars, where strong radiation fields drive significant departures from LTE in ionization and excitation states \citep{2015tsaa.book.....H, 2019A&A...621A..85H}.
        
        To properly account for these non-LTE effects, TLUSTY uses opacity sampling and extensive atomic data. This combination facilitates rigorous modeling of the radiation field, particularly in the ultraviolet and optical wavelengths. Additionally, the code accounts for line blanketing using frequency-dependent opacity sampling rather than prebinned ODFs. This method captures the cumulative opacity from metallic lines -- especially those of iron-peak elements -- which strongly modify the temperature structure and emergent energy distribution \citep{1995ApJ...439..875H}.

        The OSTAR2002 \citep{2003ApJS..146..417L} and BSTAR2006 \citep{2007ApJS..169...83L} grids, hereafter OSTAR and BSTAR, provide the standard TLUSTY-based models for hot, massive stars. The OSTAR grid spans from $\Teff = 27500\,\K$ to $55000\,\K$, while the BSTAR grid covers the $15000 - 30000\,\K$ range. Both grids extend to a maximum surface gravity of $\logg = 4.75$, and the lowest value is temperature-dependent; it corresponds to the hydrostatic stability limit near the Eddington luminosity. These TLUSTY models are computed with steps of $0.25\,\dex$ in $\logg$ and with temperature intervals of $\Delta \Teff \left[\K\right] = 2500$, $1000$ for the OSTAR and BSTAR grids, respectively (see Table~\ref{T: grids_params} and Fig.~\ref{Fig: grids_parameters}). In terms of chemical composition, both grids encompass eight metallicities between $Z/\Zsol=0$ and $2$ \citep{2025AJ....169..178H}\footnote{Adopted solar composition from \citet{1998SSRv...85..161G}.}.

        The BSTAR and OSTAR grids overlap within the $27500 \leq \Teff \left[\K\right]\leq 30000$ interval, although their synthetic spectra are not identical. These discrepancies arise from differences in chemical compositions and ion sets: the OSTAR models include higher ionization stages, whereas the BSTAR grid incorporates lower ions and neutral species \citep{2003ApJS..146..417L, 2007ApJS..169...83L}. Because the OSTAR models provide a more physically consistent description in this overlapping regime \citep{2025AJ....169..178H}, we adopted the OSTAR grid for all computations within the $27500-30000\,\K$ range.
        }
    }

\section{The \PRISMAS{} pipeline: Spectral synthesis and intensity-grid generation for \SPAMMS{}} \label{S: PRISMAS}
    {
        {
        The \SPAMMS{} code describes the local radiation field at the stellar surface using specific intensities, $I(\lambda, \mu \equiv \cos{\theta})$. Here, $\lambda$ denotes the wavelength, and $\theta$ is the angle between the surface normal and the line of sight.
        
        $\Ilambmu$ is obtained by solving the RTE. Its solution depends directly on the adopted stellar model atmosphere, which defines the depth-dependent thermal and density structure of the star -- in particular, temperature, gas pressure, and electron density as a function of the optical depth. Therefore, model atmospheres establish the local physical conditions for absorption, emission, and scattering processes. 

        In this work, we computed $\Ilambmu$ by combining the spectral synthesis capabilities of \synple{} with the automated grid-generation framework of \PRISMAS{}. Specifically, \synple{} performs the radiative transfer calculations from precomputed model atmospheres, obtaining specific intensities. At the same time, \PRISMAS{} manages the large-scale computation of stellar grids. Hence, the integration of both tools provides a fully automated and efficient workflow for producing extensive intensity grids.
        }

    \subsection{Radiative transfer and spectral synthesis with \synple} \label{S: synple}
        {
        We calculated $\Ilambmu$ using \synple{} \citep{2021arXiv210402829H}. This ``diagnostic'' code acts as a wrapper for the Fortran code \SYNSPEC{}, which efficiently solves the RTE for multiple model atmosphere frameworks \citep{2017arXiv170601859H, 2021arXiv210402829H}.

        \SYNSPEC{} calculates the formal solution using a method that depends on the input model atmosphere \citep{2021arXiv210402829H}. Computations with Kurucz models require only the physical structure and chemical compositions as a function of depth. The code first determines atomic and molecular occupation numbers under LTE, then calculates the opacities and emissivities before solving the RTE. In contrast, for TLUSTY, \SYNSPEC{} evaluates the formal integral directly. That is, it uses the atmospheric structure and the precomputed occupation numbers from the original non-LTE solution.
                  
        Nevertheless, both approaches rely on accurate opacities and emissivities to solve the RTE. \SYNSPEC{} computes them using comprehensive atomic and molecular data, including photoionization cross sections, transition probabilities, and damping constants. We therefore incorporated these datasets into \synple{} after careful review and validation for both the Kurucz and TLUSTY models:
\newpage
        \begin{enumerate}[label=(\roman*)]
            \item For the Kurucz grids, we adopted the latest atomic and molecular parameters from \citet{2018A&A...618A..25A}. In particular, the dataset includes updated ExoMol data for several molecules, such as titanium oxide and water \citep{2012MNRAS.425...21T}. Consequently, this improvement significantly enhances the accuracy of molecular opacities, which are critical for modeling cool stars. \\
            
            \item For TLUSTY, we used the original line list and atomic datasets from the OSTAR and BSTAR grids \citep{2003ApJS..146..417L, 2007ApJS..169...83L}. Our treatment of iron lines, however, differed from the initial grids: we did not use the superline approximation directly. Instead, we first computed the LTE population for each superline and compared it with its non-LTE counterpart. Then, we applied the resulting correction factor to the LTE population of each individual transition within that superline. This procedure yields NLTE-corrected populations for individual lines, relaxing the constraints of the superline approximation (see \citealt{2018A&A...618A..25A}, where the same approach was adopted).
        \end{enumerate}
        }

    \subsection{Grid generation for \SPAMMS{}: The \PRISMAS{} pipeline} \label{S: grid_generation}
        {
        To compute the intensity grids, we developed \PRISMAS{} (Pipeline of Radiative Intensity Synthesis for Meshed Atmospheric Surfaces). This script acts as a dedicated wrapper for \synple, and it is designed to bridge the gap between standard 1D model atmospheres and the \SPAMMS{} framework.

        \PRISMAS{} handles the computation of specific intensities and calibrated fluxes, with configurable wavelength ranges, spectral sampling ($\Delta\lambda$), and microturbulent velocities. \PRISMAS{} also includes automated parallel execution and data-management routines optimized for large grid computations \citep[e.g., HTCondor;][]{2024zndo..11397217H}. Finally, \PRISMAS{} organizes, compresses, and stores the computed data in a homogeneous format designed for the \SPAMMS{} framework.
        
        In this work, we employed \PRISMAS{} to generate intensity grids over the ultraviolet and optical range ($3000-9000\,\AA$), using the Kurucz and TLUSTY model atmospheres of Sect.~\ref{S: stellar_models}. Our computations adopted four discrete microturbulences ($1$,~$3$,~$5$,~$10\,\kmps$)\footnote{The input atmosphere structures of Kurucz incorporate a native microturbulence of $\xi = 2\,\kmps$. Similarly, the TLUSTY-based OSTAR and BSTAR grids adopt initial $\xi$ values of $10\,\kmps$ and $2\,\kmps$, respectively. \synple{} incorporates microturbulence during the computation of the line absorption profiles, before solving the RTE.} and a fixed spectral resolution with $\Delta\lambda = 0.01\,\AA$.

        The original implementation of \synple{} computes specific intensities at only ten values of $\mu$. For \PRISMAS{}, we modified \synple{} to compute $\Ilambmu$ on a refined grid of $101$ uniformly spaced $\mu$ values. This increased angular resolution provides a more accurate representation of the radiation field, as shown by the convergence tests of Appendix~\ref{A: mu_convergence}.

        In our intensity grids, reference abundances are not standardized across the Kurucz and TLUSTY models. \synple{} preserves the solar composition adopted in each atmospheric grid, ensuring internal consistency with the underlying model atmosphere. Consequently, spectra computed with identical stellar parameters but from different grids do not share the same chemical abundances. The maximum discrepancy reaches approximately $+0.2\,\dex$ for carbon and oxygen, while remaining elements exhibit smaller composition offsets.

        Moreover, the adopted definition of metallicity is not uniform across the computed grids. For TLUSTY, we adopted the metallicity parameter $Z$, while for Kurucz we used the logarithmic metallicity $\MH$. This choice follows the conventions of the respective model atmospheres and preserves their original parameterization. For reference, $\MH$ can be defined relative to $Z$ as
        \begin{equation}
            \MH =
                \log_{10}\left({\frac{Z}{\Zsol}}\right).
            \label{Eq: M/H}
        \end{equation}

        In addition to specific intensities, we computed Eddington fluxes and their continuum levels as
        \begin{equation}
            H\left(\lambda\right) =
                        \frac{1}{2}\int_0^1 I\left(\lambda, \mu\right)\mu \, {\rm d}\mu,
            \label{Eq: Edd_flux}
        \end{equation}
        which assumes spherical geometry and a uniform effective temperature and surface gravity\footnote{Eddington fluxes are not required by \SPAMMS{}. However, they are included in the computed grids for completeness and potential use in other applications.}. The resulting fluxes are expressed in $\mathrm{erg\,s^{-1}cm^{-2}\AA^{-1}}$, consistent with the units adopted for the specific intensities.

        Ultimately, we generated a comprehensive dataset of intensity grids for \SPAMMS{}. The new grids cover the ultraviolet-optical range, under both LTE and non-LTE assumptions. The parameter space spans effective temperatures between $\Teff = 3500\,\K$ and $55000\,\K$, surface gravities from $\logg=0.0$ to $\mathrm{5.0}$, and includes multiple metallicities ($Z/\Zsol = 0-30$) and microturbulent velocities ($\xi = 1-10\,\kmps$). As a result, \SPAMMS{} can now model nonspherical systems across the entire Hertzsprung-Russell diagram.
        }

    \subsection{Grid format, data access, and implementation into \SPAMMS{}} \label{S: grid_format}
        {
        Our intensity grids are organized in a hierarchical directory structure, where folder names encode the adopted atmospheric grid and the corresponding stellar parameters ($T_{\rm eff}$, $\logg$, metallicity, and microturbulence). Each model contains the wavelength vector, the specific intensity matrix, and the flux array, together with their respective continuum levels. To ensure compatibility with \SPAMMS{}, the intensity matrices are stored as 2D NumPy arrays: the first axis represents the $\mu$ values, while the second corresponds to the wavelength.

        The grids are not publicly hosted because of their large storage requirements. Instead, they can be obtained by contacting the authors directly. To facilitate their use, each dataset is provided together with the documentation required for its integration into the \SPAMMS{} framework.

        To ensure consistency within each modeling run, \SPAMMS{} is configured to use a single intensity grid. This choice reflects the fact that the different grids are based on distinct physical assumptions and numerical implementations -- including the LTE treatment, opacity sources, atomic data, line lists, and metallicity parameterization. As a result, $\Ilambmu$ computed from different model atmospheres are not necessarily physically equivalent, even when the same stellar parameters are adopted. Therefore, combining intensity grids from different model atmospheres within a single \SPAMMS{} run would introduce systematic inconsistencies.

        \begin{figure*}[th!]
            \centering
            \resizebox{\hsize}{!}{\includegraphics{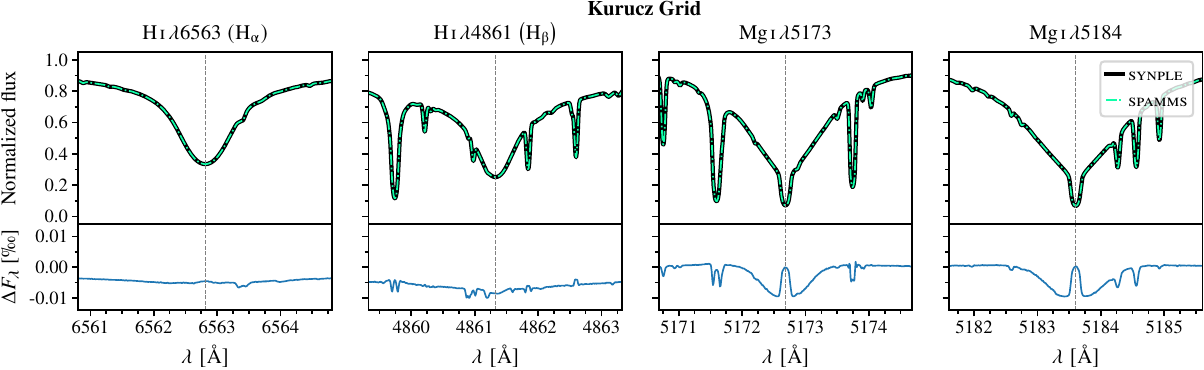}}\\[1ex]
            \resizebox{\hsize}{!}{\includegraphics{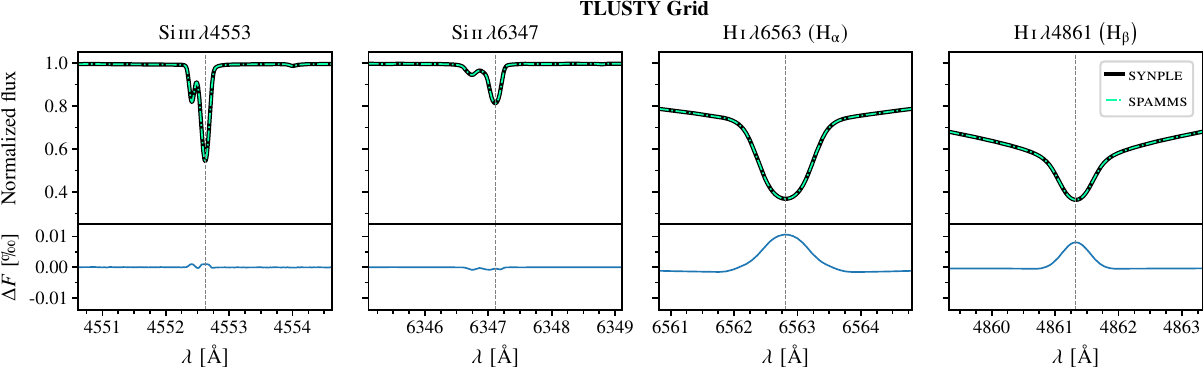}}
            \caption{
               Comparison of synthetic spectra computed with \SPAMMS{} and \PRISMAS{} for spherically symmetric, nonrotating stars. Top: Results for a solar-type star ($\Teff = 5750\,\K$ and $\logg = 4.50$), based on the ATLAS9-Castelli\&Kurucz models \citep{2003IAUS..210P.A20C}. Bottom: Spectra for a B-type star ($\Teff = 21000\,\K$, $\logg = 4.00$, $\Rpole = 4.5\,\Rsol$, and $\Mstar = 7\,\Msol$), computed using the BSTAR grid \citep{2007ApJS..169...83L}. Each subplot compares the normalized fluxes in the top panel. The bottom panel shows the flux differences, calculated according to Eq.~\ref{Eq: flux_diff} and expressed in parts per thousand (\textperthousand). The selected wavelength intervals contain key diagnostic lines for spectral analysis. The vertical dashed lines indicate their central wavelengths.
            }
            \label{Fig: spherical}
        \end{figure*}  
        }
    }

\section{Numerical validation of the surface integration scheme in \SPAMMS{}} \label{S: validation}
    {
    The \SPAMMS{} code characterizes the local radiation field using specific intensities and computes the stellar flux as
    \begin{equation}
        F_\lambda =\sum_n \Ilambmu_n a_n \, \mu_n \, v_n.
        \label{Eq: spamms_integration}
    \end{equation}
    Here, the subscript $n$ refers to a triangle in the surface mesh, $\Ilambmu_n$ is the emergent specific intensity at wavelength $\lambda$, $a_n$ the triangle area, $\mu_n$ the cosine of the emergent angle, and $v_n$ is a visibility factor equal to 1 for visible triangles and 0 otherwise \citep{2020A&A...636A..59A}.

    While the surface integration scheme in \SPAMMS{} remains unchanged, its application has been extended. \SPAMMS{} was originally developed and validated using \FASTWIND{}, which provides isolated line profiles. In this work, the same scheme is applied to Kurucz and TLUSTY grids, which output full spectra -- including both the continuum and spectral lines. The purpose of this section is therefore not to validate a new integration method, but to verify that the existing scheme performs consistently when applied to full-spectrum inputs.

    To this end, we compared \SPAMMS{} and \PRISMAS{} spectra from spherical, nonrotating stars with identical stellar parameters. As a result, we isolated the numerical effects of the surface integration process: any detected flux difference originates from the integration method rather than the input model atmosphere. 
    
    For Kurucz, we adopted a solar-type model from the ATLAS9-Castelli\&Kurucz grid with $\Teff = 5750\,\K$, $\logg = 4.50$, and a microturbulence of $\xi = 1\,\kmps$. In practice, since \PRISMAS{} and \SPAMMS{} follow the same computational procedure for all Kurucz models, the results are directly applicable to the M{\'e}sz{\'a}ros grid. For the TLUSTY validation, we adopted a B-type model from the BSTAR grid. We set $\Teff = 21000\,\K$, $\logg = 4.00$, $Z = \Zsol$, and $\xi = 1\,\kmps$, with a polar radius $\Rpole = 4.5\,\Rsol$ and a stellar mass $\Mstar = 7\,\Msol$.

    Figure~\ref{Fig: spherical} compares the synthetic spectra computed with \SPAMMS{} and \PRISMAS{} around key diagnostic lines. The upper panels show the normalized spectra, while the lower ones display their relative flux differences, defined as
    \begin{equation}
        \Delta F_\lambda =
                        \frac{F_j(\lambda)-F_i(\lambda)}{F_i(\lambda)},
        \label{Eq: flux_diff}
    \end{equation}
    with $F_i = F_\PRISMAS{}$ and $F_j = F_\SPAMMS{}$. For the B-type star, we examined the \specline{Si}{ii}[6347] and \specline{Si}{iii}[4553] lines, which are standard effective temperature diagnostics. We also included Balmer lines (\specline{H}{i}[4861], \specline{H}{i}[6563]), commonly used to constrain the surface gravity \citep{2024A&A...687A.228D}. For the solar-type model, we considered the \specline{Mg}{i} triplet at $5173\,\AA$ and $5184\,\AA$, which is sensitive to gravity in cool stellar atmospheres. We also displayed two Balmer lines used for temperature validation in this regime \citep{2008oasp.book.....G}.
    
    The comparison reveals excellent agreement between both synthesis methods, with flux deviations below $0.01\text{\textperthousand}$ across all diagnostic lines. Specifically, these negligible discrepancies stem from differences in how the integration is performed. In \SPAMMS{}, the stellar surface is discretized into a mesh, and the total flux results from the integration of emergent intensities across the individual patches (Eq.~\ref{Eq: spamms_integration}). Thus, this procedure introduces minor numerical approximations associated with the finite surface resolution. In contrast, \PRISMAS{} directly computes the emergent flux from the specific intensities (Eq.~\ref{Eq: Edd_flux}), which minimizes numerical noise. However, this method is restricted to perfectly symmetric atmospheres and cannot account for deviations from spherical geometry. Therefore, \SPAMMS{} accurately reproduces the fluxes predicted by 1D codes, while providing the capability to model different stellar shapes.
    }


\section{Applications of \SPAMMS{} in 3D stellar modeling} \label{S: applications}
    {    
        {
        \begin{figure*}[th!]
            \centering
            \resizebox{\hsize}{!}{\includegraphics{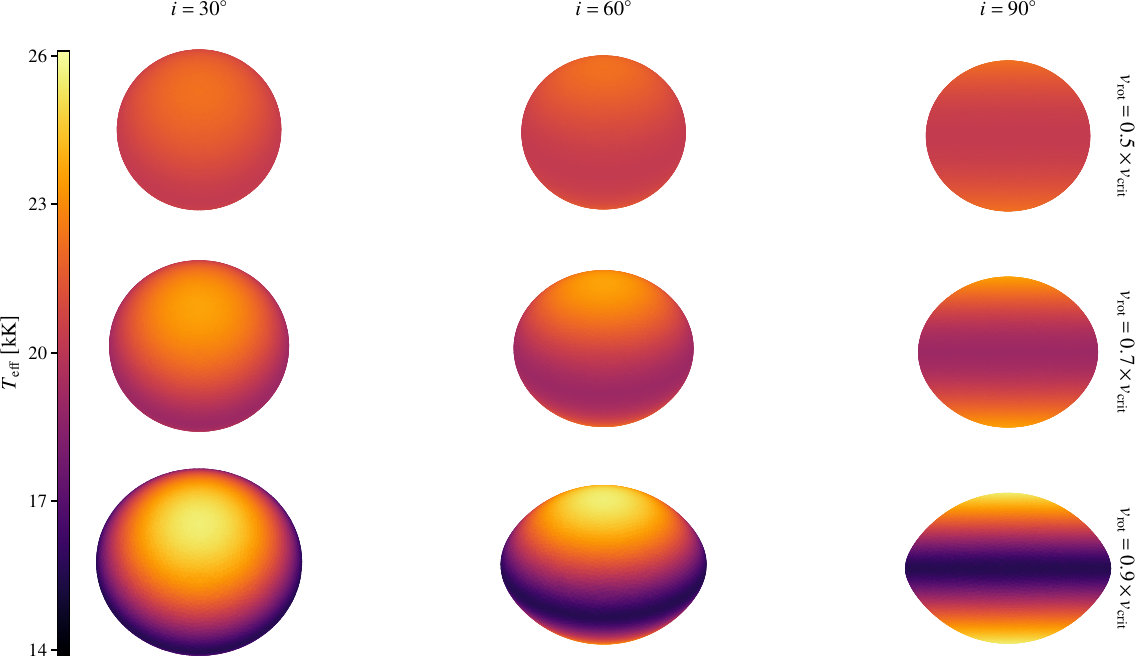}}
            \caption{
                Effective temperature distribution across the surface for a rotating B-type star with $\Teff = 21000\,\K$, $\Mstar = 7\,\Msol$, and $\Rpole = 4.5\,\Rsol$. The vertical color bar indicates the local surface temperature: darker and lighter colors represent lower and higher values, respectively. The rows indicate rotation rates at $50\%$, $70\%$, and $90\%$ of the critical equatorial velocity, $\vrotcrit$. Each column corresponds to a different inclination angle, $i$, of the rotation axis ($30^\circ$, $60^\circ$, and $90^\circ$).
            }
            \label{Fig: rotatingBstar_Teff}
        \end{figure*}
        
        Conventional stellar atmosphere codes generally assume spherical symmetry (see \citealt{2024arXiv240903329P} for a review and further references). However, this assumption often fails for real stars: rapid rotation and binary interactions produce strong departures from spherical geometry. In fact, these distortions induce local surface variations in atmospheric parameters, which modify the emergent radiation field and the resulting spectrum.

        Rapid rotation produces an oblate distortion through centrifugal forces, causing an equatorial bulge relative to the polar regions \citep{2009pfer.book.....M}. The resulting deformation generates a latitudinal variation in the surface gravity, $\geff$, with poles showing higher values than the equator \citep{1924MNRAS..84..665V}. Because $\Teff$ scales directly with $\geff$ \citep{1999A&A...347..185M}, this gravity distribution induces a latitudinal thermal gradient (with hotter poles and cooler equatorial regions), which intensifies with increasing rotation and significantly modifies the emergent radiation field. As a result, both the spectral energy distribution and line profiles critically dependent on the rotation rate and inclination of the rotation axis \citep{2023A&A...669L..11A}.
    
        Similarly, stellar multiplicity alters stellar structure, the emergent radiation field, and the evolutionary pathways. Tidal interactions distort the stellar surface, producing spatial variations in $\geff$ and $\Teff$ \citep{2018maeb.book.....P}. Binary interactions also enable mass transfer, angular momentum exchange, and spin synchronization, which have a deep impact on stellar evolution \citep{2012ARA&A..50..107L}. Moreover, radiative reflection causes radiation from the primary component to heat the facing hemisphere of the companion \citep{2016ApJS..227...29P,2016AAS...22734418G}. Consequently, these interactions alter the surface temperature distribution, modify the observed spectral line strengths, and ultimately change the observed brightness of the system.

        We present two representative test cases to demonstrate the capability of \SPAMMS{} to model 3D stellar surfaces: (i) a rapidly rotating B-type star, which exemplifies centrifugal distortion and (ii) an eclipsing Algol-type binary, showing tidal deformation and mutual irradiation. 
        }

    \subsection{Rotating B-type star} \label{S: rotatingBstar}
        {
        We computed a set of \SPAMMS{} models for a B-type star of effective temperature $\Teff = 21000\,\K$, stellar mass $\Mstar = 7\,\Msol$, and polar radius $\Rpole = 4.5\,\Rsol$. We adopted the BSTAR grid with solar metallicity ($Z/\Zsol = 1$) and a microturbulence of $1\,\kmps$, using the gravity-darkening prescription from \citet{2011A&A...533A..43E}.

        We parameterized stellar rotation using the linear rotation rate, which is defined as
        \vspace{0.2cm}
        \begin{equation}
            \omega =
                    \frac{\vrot}{\vrotcrit}.
        \end{equation}
        Here, $\vrot$ is the linear velocity at the equator and $\vrotcrit$ is the critical equatorial velocity, given by
        \vspace{0.2cm}
        \begin{equation}
            \vrotcrit =
                        \sqrt{\frac{G \Mstar}{1.5\Rpole}}.
        \end{equation}
        This velocity represents the physical limit of hydrostatic stability, where centrifugal and gravitational accelerations balance at the equator. Above this threshold, the equatorial surface becomes gravitationally unbound. In our analysis, we considered three rotation rates ($\vrotrate~=~0.5$,~$0.7$,~$0.9$) and three inclination angles ($30^\circ$,~$60^\circ$,~$90^\circ$) to explore the effects of rapid rotation.

        Figure~\ref{Fig: rotatingBstar_Teff} shows the surface effective temperature distribution for the B-type star, where centrifugal expansion reduces the $\Teff$ toward the equator -- producing a latitudinal temperature gradient. At low rotation rates ($0.5 \times \vrotcrit$), the temperature difference between the poles and the equator is only $\sim 2\,\mathrm{k}\K$. However, the contrast becomes increasingly pronounced with higher rotation: at $0.9\times\vrotcrit$, the equator reaches approximately $16\,\mathrm{k}\K$, while the poles have $\Teff \sim 22\,\mathrm{k}\K$.

        The apparent temperature of the star also depends on inclination, as the viewing angle sets which latitudes contribute most to the observed flux (Fig.~\ref{Fig: rotatingBstar_Teff}). Hotter polar regions dominate the observed surface at low inclinations, resulting in a higher apparent temperature and a bluer integrated spectrum. Conversely, an edge-on view highlights the cooler equator, reducing the apparent temperature. At intermediate inclinations, polar and equatorial contributions are comparable.

        The surface effective gravity distribution mirrors the effective temperature variations described above. \citet{1924MNRAS..84..665V} showed that, for radiative envelopes,
        \vspace{0.05cm}
        \begin{equation}
            \Teff\left(\omega, \theta\right) \propto 
                                            \geff\left(\omega, \theta\right)^{0.25},
        \end{equation}
        \vspace{0.05cm}
        \noindent where $\theta$ denotes the local colatitude, and $\geff\left(\theta\right)$ is the effective gravity including centrifugal acceleration. \citet{2011A&A...533A..43E} refined this scaling by introducing a gravity-darkening factor, $\Xi\left(\theta,\omega\right)$, that depends on colatitude and rotation rate:
        \vspace{0.05cm}
        \begin{equation}
             \Teff\left(\omega, \theta\right) \propto 
                                                \Xi\left(\omega,\theta\right)\geff\left(\omega, \theta\right)^{0.25}.
        \end{equation}
        \vspace{0.05cm}
        \noindent Following this relation, the equatorial regions exhibit lower $\logg$ values, while the poles reach higher ones (see Fig.~\ref{Fig: rotatingBstar_loggeff} in Appendix~\ref{A: rotatingBstar}). As a result, the apparent surface effective gravity also varies with the inclination angle: pole-on views emphasize high-$\logg$ regions, whereas equator-on orientations highlight the lower-$\logg$ zones.

        \begin{figure}[th!]
            \centering
            \resizebox{\hsize}{!}{\includegraphics[scale=1]{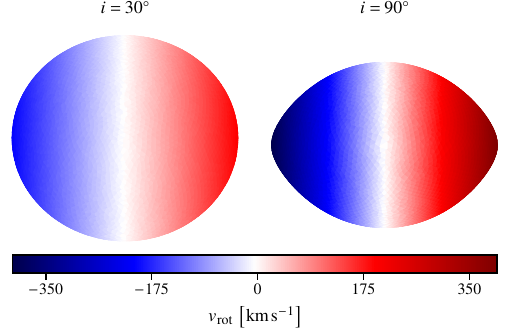}}
            \caption{
                Radial velocity distribution of a rotating B-type star, with $\Teff = 21000\,\K$, $\Mstar = 7\,\Msol$, $\Rpole = 4.5\,\Rsol$, and a linear velocity at the equator of $\vrot \sim 400 \,\kmps = 0.9 \times \vrotcrit$. The panels display different inclination angles of the rotation axis ($i = 30^\circ, 90^\circ$). The color bar indicates the local line-of-sight velocity component: bluer and redder colors denote regions approaching and receding from the observer, respectively.
           }
            \label{Fig: rotatingBstar_rot}
        \end{figure} 

        Moreover, measuring stellar rotation is particularly challenging because spectroscopy constrains only the projected rotational velocity, $\vsini$, through line broadening. Therefore, the measured velocity depends strongly on the inclination angle: pole-on stars exhibit narrow spectral lines and low $\vsini$ values, while equator-on views display the largest projected velocities (Fig.~\ref{Fig: rotatingBstar_rot}). Thus, neglecting the inclination effect introduces systematic errors, affecting not only the inferred rotation rate but also other fundamental parameters -- such as effective temperature and abundances.

        Finally, we compared spectra computed with two different rotation treatments in Fig.~\ref{Fig: rotatingBstar_spec}. The first method models a distorted stellar surface with \SPAMMS{}, as shown in Fig.~\ref{Fig: rotatingBstar_Teff}. The second approach convolves the intrinsic, nonrotating \SPAMMS{} spectra with the Doppler-shift kernel of \citet{1933MNRAS..93..478C}. To facilitate comparison, we computed the flux differences using Eq.~\ref{Eq: flux_diff}, where $F_i$ and $F_j$ represent the \SPAMMS{} and convolved models, respectively. 
        
        In particular, our analysis focused on the \specline{He}{i}[4471] line and the \specline{Mg}{ii}[4481] triplet, which are highly sensitive to effective temperature \citep{2008oasp.book.....G}. \specline{He}{i}[4471] includes a forbidden component at $\lambda4470$, which blends with the permitted component at high rotation rates -- producing an asymmetric blueshifted profile. \specline{Mg}{ii}[4481] also presents a weaker but persistent asymmetry, arising from the intrinsic lines at $4481.13\,\AA$ and $4481.33\,\AA$.
        
        Discrepancies between the distorted surface model and the convolution approach are negligible at low rotation rates ($<~2\%)$. With increasing rotational velocity, the \specline{He}{i}[4471] profiles exhibit progressively larger deviations, reaching maximum differences of $4\%$. Thus, the convolution overestimates the \specline{He}{i}[4471] line depth, mimicking lower effective temperatures or an enhanced helium abundance. Discrepancies in the \specline{He}{i} profiles are also stronger at higher inclinations, where the observed surface is dominated by cold, rapidly rotating equatorial regions. 
        
        In contrast, the \specline{Mg}{ii}[4481] profile remains nearly unchanged in both modeling treatments, with flux differences below $1\%$. Yet, at the highest rotation rates and inclinations, its blue wing flattens only for the convolution approach. Therefore, this profile distortion causes the line to be difficult to fit when comparing convolved synthetic and observed spectra.

        \begin{figure*}[h!]
            \centering
            \resizebox{\hsize}{!}{\includegraphics{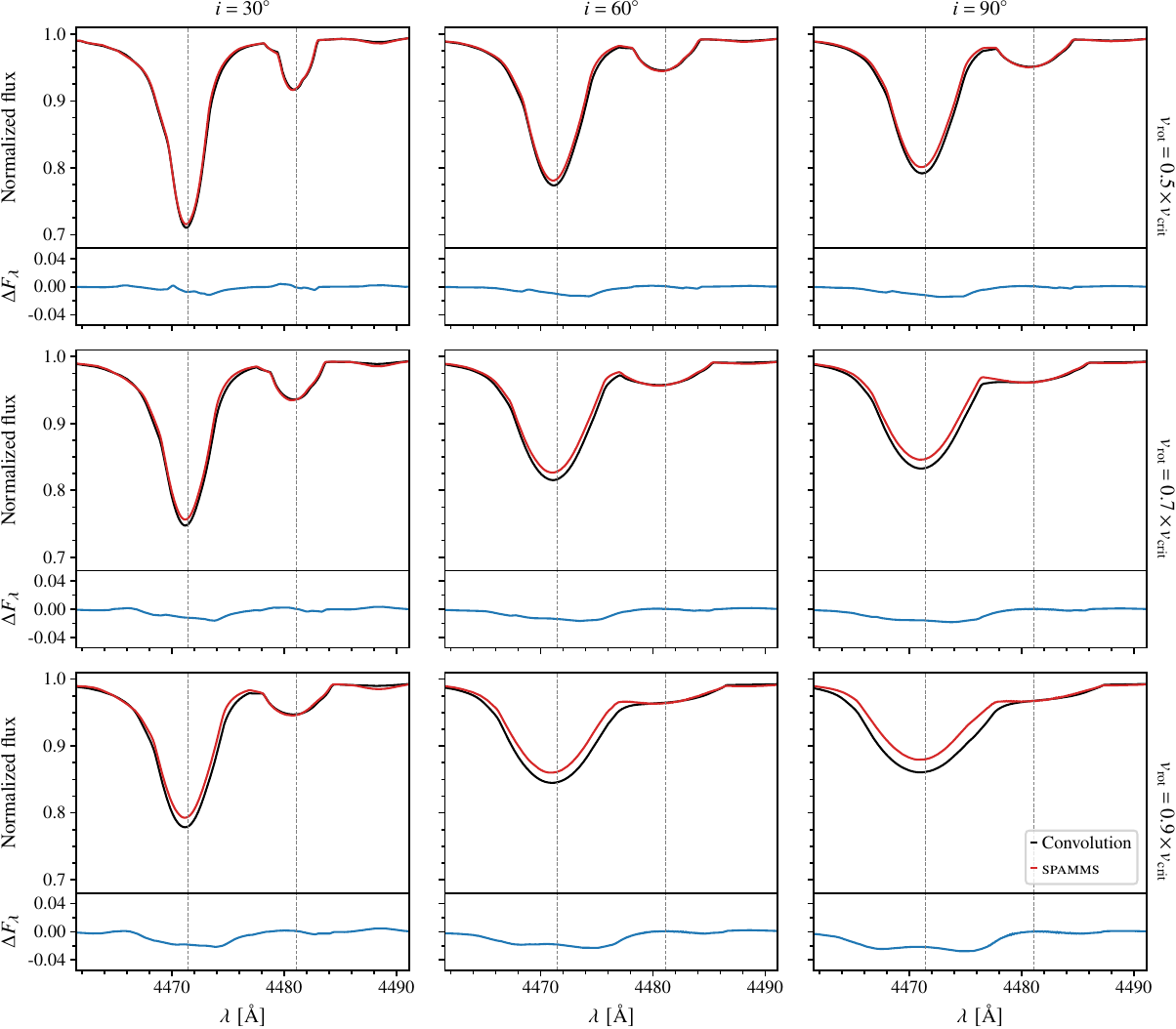}}
            \caption{
                Comparison of spectra for a rotating B-type star ($\Teff = 21000\,\K$, $\Mstar = 7\,\Msol$, and $\Rpole = 4.5\,\Rsol$), computed using the BSTAR grid \citep{2007ApJS..169...83L}. Top panels: \SPAMMS{} spectra (red profiles) explicitly accounting for the effects of rotational geometry (see Figs.~\ref{Fig: rotatingBstar_Teff} and \ref{Fig: rotatingBstar_loggeff}). The black profiles represent classical broadening, obtained by convolving the intrinsic, nonrotating \SPAMMS{} spectrum with a Doppler-shift kernel \citep{1933MNRAS..93..478C}. Bottom panels: Flux differences, calculated according to Eq.~\ref{Eq: flux_diff}, with $F_i$ and $F_j$ corresponding to the rotating-\SPAMMS{} and convolved models, respectively. Rows: Rotation rates at $50\%$, $70\%$, and $90\%$ of the critical equatorial velocity, $\vrotcrit$. Columns: Inclination angles of the rotation axis ($i = 30^\circ$,~$60^\circ$,~$90^\circ$). The vertical dashed lines indicate the \specline{He}{i}[4471] line and \specline{Mg}{ii}[4481] triplet at their rest wavelengths.
            }
            \label{Fig: rotatingBstar_spec}
        \end{figure*}

        To summarize, the \ratio{He}{i}{Mg}{ii} ratio decreases with increasing rotation rate and inclination. However, although this trend is reproduced by both rotation treatments, the inferred effective temperatures depend on the adopted model. In particular, the convolution approach systematically gives lower effective temperatures because it overestimates the \specline{He}{i}[4471] line depth and flattens \specline{Mg}{ii}[4481].

        Consequently, the results of this section show that broadening kernels fail to accurately capture the effects of rotational geometry; these models account for neither temperature and gravity gradients nor surface deformation, making the spectral analysis of fast rotators unreliable. Hence, only full surface integration -- as implemented in \SPAMMS{} -- can accurately reproduce line profiles and enable the reliable determination of stellar parameters.
        }

    \subsection{Eclipsing Algol-type binary} \label{S: algol}
        {
        \begin{figure*}[th!]
            \centering
            \resizebox{\hsize}{!}{\includegraphics{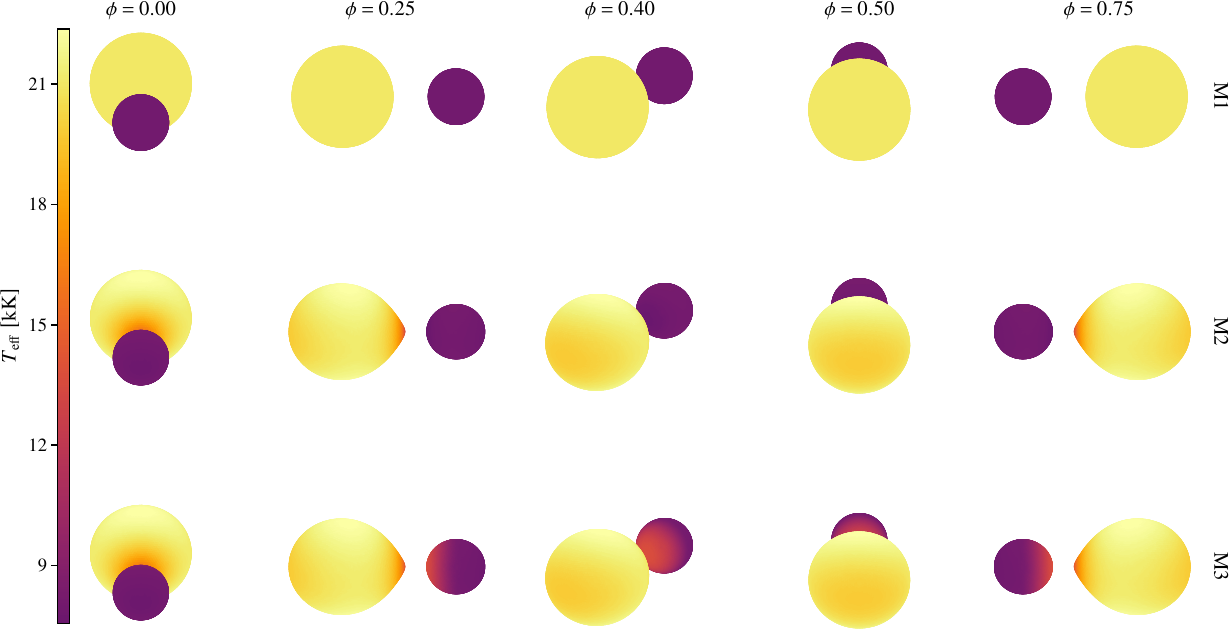}}
            \caption{
                \SPAMMS{} surface modeling of an eclipsing Algol-type binary at five key orbital phases. The adopted parameters correspond to a hot \spectype{B}{2}[IV] primary and a cooler \spectype{A}{7}[V] secondary (Table~\ref{T: algol_parameters}). The different panels compare three modeling configurations to isolate geometric and radiative effects. (i) Top row: Spherical geometry excluding radiative reflection. (ii) Middle row: Roche geometry without radiative reflection. (iii) Bottom row: Roche geometry and radiative reflection.
            }
            \label{Fig: algol_modeling}
        \end{figure*}
        
        Algol-type eclipsing binaries are close systems in which the initially more massive star has evolved to fill its Roche lobe. In these systems, strong interactions -- such as tidal distortion, surface heating, and mass transfer -- significantly modify the stellar structure and emergent radiation. 
        
        For this section, we modeled an Algol binary consisting of a hot \spectype{B}{2}[IV] primary and a cooler \spectype{A}{7}[V] secondary, with the orbital and physical parameters of Table~\ref{T: algol_parameters}. In this specific configuration, the primary is an evolved subgiant undergoing Roche-lobe overflow, with the system in the pre-mass-transfer phase. 

        \begin{table}[h!]
            \caption{
                Main physical and orbital parameters of the Algol system.
            } 
            \label{T: algol_parameters}
            \centering   
            \small
            \begin{tabular}{ccc}
                \hline
                \hline \\ [-1.8ex]
                Parameters         & Primary     & Secondary    \\ \hline \\ [-1.8ex]
                $P$ [$\mathrm{d}$] & \multicolumn{2}{c}{$1.6$}  \\ [0.3ex]
                $q$                & \multicolumn{2}{c}{$0.5$}  \\ [0.3ex]
                $a$ [$\Rsol$]      & \multicolumn{2}{c}{$12.5$} \\ [0.3ex]
                $i$ [deg]          & \multicolumn{2}{c}{$70.0$} \\ [0.3ex]
                $e$                & \multicolumn{2}{c}{$1.0$}  \\ [0.3ex]
                $\Mstar$ [$\Msol$] & $6.0$       & $3.0$        \\ [0.3ex]
                $\Reqv$ [$\Rsol$]  & $5.5$       & $3.0$        \\ [0.3ex]
                $\Teff$ [$\K$]     & $21000$     & $8000$       \\ [0.3ex]
                $F$                & 1           & 1            \\ [0.3ex]
                $\beta$ [$\deg$]   & 0           & 0            \\ [0.3ex]
                $\lambda$ [$\deg$] & 0           & 0            \\ \hline
            \end{tabular}
            \tablefoot{
                $P$: orbital period; $q = M_2/M_1$: mass ratio; $a$: semimajor axis; $i$: orbital inclination; $e$: eccentricity; $\Mstar$: stellar mass; $\Reqv$: equivalent radius; $\Teff$: effective temperature; $F = \Omega_\mathrm{rot}/\Omega_\mathrm{orb}$: synchronicity ratio; $\beta$: pitch angle; $\lambda$: yaw angle.
                }
        \end{table}

        Figure~\ref{Fig: algol_modeling} illustrates the \SPAMMS{} surface modeling at five orbital phases ($\phi = 0.0$, $0.25$, $0.40$, $0.50$, and $0.75$) for three configurations:
        \begin{itemize}
            \item M1: Spherical geometry without radiative reflection;
            \item M2: Roche geometry with no radiative reflection;
            \item M3: Roche geometry with radiative reflection enabled.
        \end{itemize} 
        Model M1 (top row) represents the idealized spherical case, in which the effective temperature is uniform across the entire stellar surface. By contrast, the middle row (M2) incorporates the Roche geometry and reveals the consequences of tidal distortion. As the primary fills its Roche lobe, tidal forces significantly elongate its stellar surface toward the companion. Consequently, the departure from spherical symmetry produces a nonuniform distribution of effective temperature over the surface. In particular, the regions associated with the tidal bulge become cooler than in the spherical model. 

        The bottom row of Fig.~\ref{Fig: algol_modeling} (model M3) shows the effect of radiative reflection, which is another critical consequence of stellar multiplicity \citep{2016ApJS..227...29P, 2016AAS...22734418G}. In this configuration, the hot primary irradiates the facing hemisphere of the cool secondary, increasing its local surface temperature and enhancing the emitted intensity from the irradiated region.

        \begin{figure*}[th!]
            \centering
            \resizebox{\hsize}{!}{\includegraphics{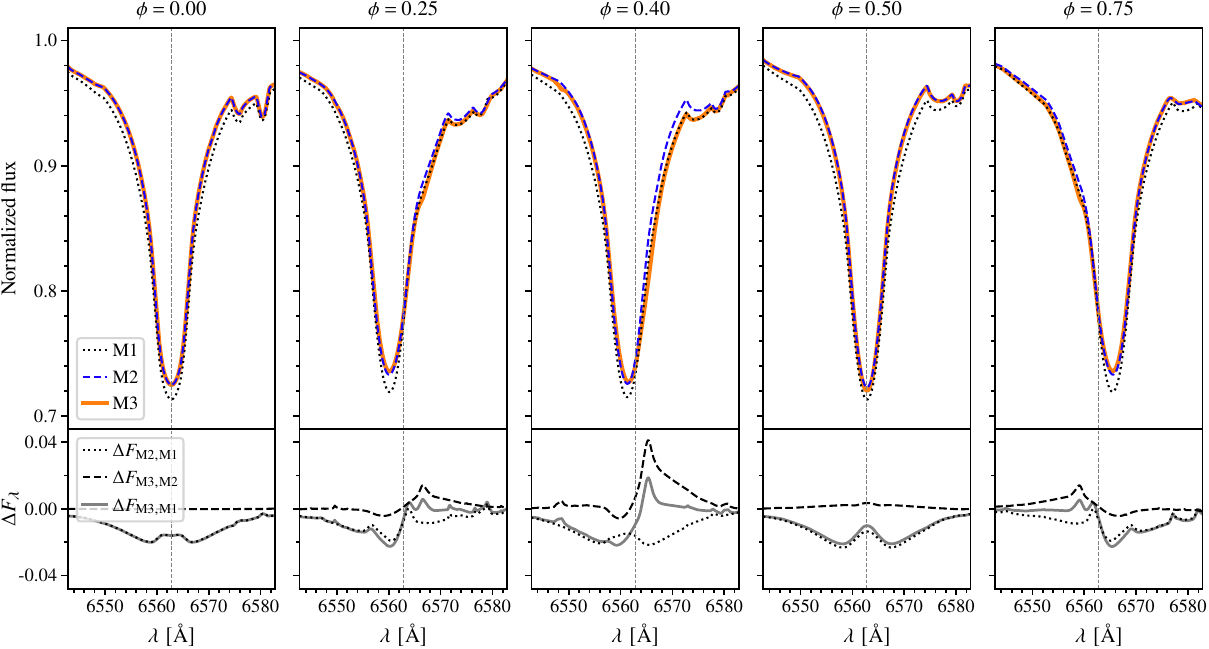}}
            \caption{
                \SPAMMS{} synthetic spectra of an Algol binary system at five orbital phases. All models use the \cite{2003IAUS..210P.A20C} grid and the parameters of Table~\ref{T: algol_parameters}. Top panels: Normalized flux for three modeling configurations. The solid orange lines denote the M3 model, which incorporates Roche geometry and radiative reflection. The dashed blue lines display the M2 model, which adopts Roche geometry without radiative reflection. The dotted black lines show the M1 model (spherical geometry with radiative reflection disabled). The vertical dashed lines indicate the $\mathrm{H_\upalpha}$ rest wavelength. Bottom panels: Flux differences, $\Delta F(\lambda)_{i,j}$, computed with Eq.~\ref{Eq: flux_diff}.
            }
            \label{Fig: algol_spec}
        \end{figure*}

        Figure~\ref{Fig: algol_spec} displays the \SPAMMS{} spectra of the Algol system, computed using the Castelli\&Kurucz grid with $\xi = 1\,\kmps$ and $\MH = 0$. We focused on the $\mathrm{H_\upalpha}$ line, as it is widely used to constrain stellar parameters -- such as the effective temperature and surface gravity \citep{2008oasp.book.....G}.

        The spectra are dominated by the $\mathrm{H_\upalpha}$ absorption line of the B-type primary, while the contribution from the cooler secondary is significantly weaker. Also, as the stars orbit each other, both spectral components undergo Doppler shifts. During the eclipses ($\phi = 0.0$ and $0.5$), the $\mathrm{H_\upalpha}$ components coincide with their rest wavelengths. By contrast, at quadrature ($\phi = 0.25$ and $0.75$), the two components reach their maximum separation, with the lines displaced toward the blue or red depending on the line-of-sight velocity.

        In addition to these radial velocity variations, the line depth changes with orbital phase. Notably, the absorption reaches a maximum during the eclipses. At these phases, the eclipsed companion contributes less to the total continuum flux, reducing spectral dilution and, consequently, increasing the apparent line strength.

        In Fig.~\ref{Fig: algol_spec}, we additionally present the flux differences computed using Eq.~\ref{Eq: flux_diff}. The quantity $\Delta F(\lambda)_\mathrm{M2,M1}$ isolates the effects of tidal distortion by comparing the Roche model without irradiation (M2) to the spherical model (M1). The largest differences occur at orbital phases $\phi = 0.0$ and $0.5$, when the tidally distorted surfaces are viewed most directly. During the primary eclipse ($\phi = 0.0$), however, the secondary partially occults the distorted hemisphere of the primary, reducing the expected maximum difference.

        On the other hand, $\Delta F(\lambda)_\mathrm{M3,M2}$ isolates the effects of radiative reflection, which is strongest outside the primary eclipse. At $\phi = 0.0$, the irradiated hemisphere of the secondary is completely hidden from the observer, and the emergent flux originates from its cooler, nonirradiated surface. As the system moves away from eclipse, the irradiated hemisphere gradually comes into view, reaching its maximum visibility around $\phi \sim 0.4$. At this phase, the reflection effect is strongest. By $\phi = 0.5$, only part of the heated hemisphere remains visible, producing comparatively smaller changes in the line profiles.

        Overall, radiative reflection modifies the line profiles by up to $\sim4\%$, while tidal distortion produces smaller variations of $\lesssim2\%$. In fact, the impact of tidal deformation increases during eclipses, and radiative reflection dominates spectral variations at quadrature. Thus, these results indicate that mutual irradiation has the larger overall impact on the synthetic spectra. Finally, $\Delta F(\lambda)_\mathrm{M3,M1}$ represents the total discrepancy between the irradiated Roche model and the nonirradiated spherical model. The maximum discrepancy is approximately $2\%$, as partial compensation between these two contributions reduces the net discrepancy.

        In Fig.~\ref{Fig: algol_LC}, we show the light curves of the Algol system. They were computed using \PHOEBE{} \citep{2016ApJS..227...29P} and the Sloan photometric system \citep{1996AJ....111.1748F, 2010AJ....139.1628D}, adopting the same physical configuration used for \SPAMMS{}.

        The synthetic light curves exhibit flux variations synchronized with the orbital period. In the simplest configuration (M1), the light curve is dominated by the eclipses, with no flux variation around quadrature. 
        
        Introducing nonspherical surfaces (M2) produces a sinusoidal modulation and deepens both eclipses. In fact, the effects of Roche geometry are most pronounced during quadrature, when the observed temperature distribution differs significantly from that of the spherical model. Consequently, the eclipse profiles become asymmetric: the ingress and egress of the secondary eclipse appear brighter than those of the primary.

        Finally, the inclusion of radiative reflection (M3) further enhances the orbital modulation. As a result, the system reaches its maximum brightness near the quadrature phases ($\phi \sim 0.25$ and $\phi \sim 0.75$), when most of the heated hemisphere is visible. In addition, radiative reflection brightens the ingress and egress of the secondary eclipse relative to M2. Thus, neglecting this effect removes important photometric signatures that constrain the properties of the secondary star -- such as its albedo and the temperature contrast with the primary.

        Taken together, these results demonstrate the importance of accounting for both geometric distortion and radiative reflection when modeling close binary systems. Despite these capabilities, \SPAMMS{} presents some limitations. \PHOEBE{} relies on static model atmospheres and therefore neglects atmospheric back-warming, which alters the temperature-pressure structure and ionization balance. Consequently, irradiation is applied by simply adjusting the local surface temperature and the corresponding emergent intensity.

        Nevertheless, \SPAMMS{} represents a significant advance over simplified binary models. While many approaches neglect tidal distortion and radiative reflection, \SPAMMS{} incorporates both effects through a full 3D treatment of the stellar surfaces. This is particularly important for close binaries, where small surface variations can introduce significant biases when neglected.

        \begin{figure}[th!]
            \centering
            \resizebox{\hsize}{!}{\includegraphics{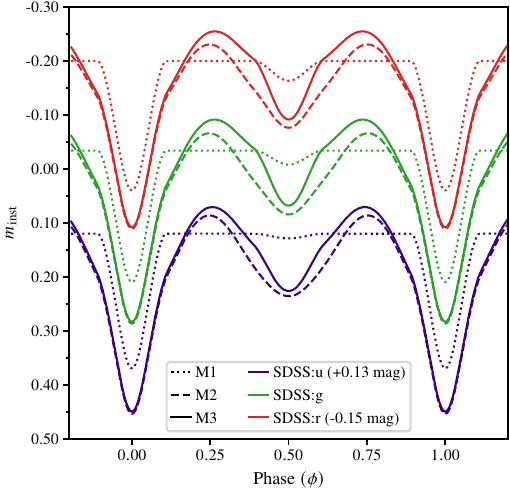}}
            \caption{
                Multiband light curves of an Algol-type binary in the Sloan \textit{u}, \textit{g}, and \textit{r} filters. The synthetic photometry was computed with \PHOEBE{} using the parameters in Table~\ref{T: algol_parameters}. The solid lines show M3, which adopts Roche geometry and radiative reflection. The dashed lines represent M2 (Roche geometry without reflection), while the dotted lines represent M1 (spherical geometry and no reflection).
            }
            \label{Fig: algol_LC}
        \end{figure}
        }
    }

\section{Conclusions} \label{S: conclusions}
    {
    We computed new intensity grids for \SPAMMS{}, a code for the 3D spectroscopic modeling of stellar surfaces \citep{2020A&A...636A..59A}. To generate these grids, we developed \PRISMAS{}, a \Python{} pipeline built around the spectral synthesis code \synple{} \citep{2021arXiv210402829H}. In addition, we used \PRISMAS{} with several precomputed model atmospheres, including two LTE ATLAS9-Kurucz grids \citep{2003IAUS..210P.A20C,2012AJ....144..120M} and the non-LTE, TLUSTY-based OSTAR2002 and BSTAR2006 datasets \citep{2003ApJS..146..417L,2007ApJS..169...83L}. The resulting grids provide specific intensities, $\Ilambmu$, over the ultraviolet-to-optical wavelength range ($3000-9000\,\AA$), covering multiple metallicities ($Z/\Zsol = 0-30$) and microturbulent velocities ($\xi = 1-10\,\kmps$).

    Moreover, we validated the \SPAMMS{} surface integration scheme against 1D flux profiles. Validation cases included a solar-type atmosphere in LTE and a B-type star in non-LTE, with relative flux deviations below $\sim 0.01\text{\textperthousand}$ between the 1D and 3D-surface models. This result demonstrates that mesh-based surface artifacts are negligible, and that the 3D integration of \SPAMMS{} successfully preserves the input model atmosphere.
    
    We further demonstrate the scientific value of the new intensity grids with two applications. The first case involves full surface modeling of a rapidly rotating B-type star, showing significant rotational and inclination effects on line profiles -- which standard convolution methods fail to reproduce. The second case examines an Algol-type binary, which exhibits tidal distortion, mutual irradiation, and orbital motion producing phase-dependent line profiles and light-curve features. These two examples demonstrate that departures from spherical symmetry require a 3D treatment of the stellar surface for accurate spectroscopic modeling.
    
    However, our intensity grids have some limitations. In particular, they do not include stellar winds, which can be significant in hot, massive stars \citep{2000ARA&A..38..613K}. Actually, previous implementations of \SPAMMS{} allowed for wind modeling, but our work prioritizes expanding the parameter space. Consequently, wind-broadened lines, P Cygni profiles, and mass-loss signatures cannot be reproduced when using \SPAMMS{} with the new grids.

    Beyond these limitations, other frameworks have recently been developed to compute the spectra of rotationally distorted stars \citep{2024A&A...685A..57L, 2024A&A...688A..97M}. However, they combine flux spectra with limb-darkening prescriptions, whereas \SPAMMS{} performs the surface integration directly using angle-dependent specific intensities. As a result, our code achieves a more self-consistent treatment of the emergent radiation field. Moreover, these new codes present other important limitations: the method of \citet{2024A&A...685A..57L} is restricted to massive stars and is not publicly available, while that of \citet{2024A&A...688A..97M} is limited to LTE model atmospheres. \SPAMMS{}, by contrast, is publicly available and is not restricted to a specific spectral type, making it a more versatile tool for the community.
    
    In conclusion, \SPAMMS{} provides a physically consistent framework that bridges 1D model atmospheres and 3D surface modeling. By combining broad coverage of stellar parameters with a detailed surface integration, \SPAMMS{} enables realistic spectroscopic modeling across the entire Hertzsprung-Russell diagram.
    }

\section{Data availability} \label{S: data}
    {
    The \PRISMAS{} pipeline is open-source and publicly available on GitHub at \url{https://github.com/DGalanDieguez/PRISMAS}. The repository includes a user guide and worked examples to facilitate its use and integration into existing workflows. 
    
    In particular, \PRISMAS{} can generate custom model grids for different combinations of microturbulent velocity, wavelength coverage, and spectral sampling. This flexibility allows the results presented in this work to be easily reproduced and extended to other configurations.
    }

    {
    \begin{acknowledgements} 
        {
        DGD, MAM, SRB, and AH acknowledge financial support from the Ministry of Science, Innovation and Universities (MCIU) through the Spanish State Research Agency (AEI), under grants PID2021-122397NB-C21 and PID2024-159329NB-C21 (co-funded by the European Regional Development Fund, FEDER). MAM is further supported by the ``la Caixa'' Foundation (ID 100010434) through fellowship LCF/BQ/PI23/11970035. SRB also benefits from the Viera y Clavijo Programme, funded by the Canary Islands Agency for Research, Innovation and Information Society (ACIISI) of the Government of the Canary Islands and the Universidad de La Laguna. CAP receives financial support from the MCIU through grants AYA2014-56359-P, AYA2017-86389-P, and PID2020-117493GB-I00. HS is supported by the Flemish Government through the long-term structural Methusalem funding programme, project SOUL: Stellar Evolution in Full Glory (grant METH/24/012), at KU Leuven. Additional funding is provided by the Research Foundation -- Flanders (FWO; grant G0ABL24N) and by the KU Leuven Research Council (grant iBOF/21/084). Moreover, the authors affiliated with the IAC recognize institutional funding from grant CEX2025-001609-S, awarded under the Severo Ochoa Centre of Excellence program (MICIU/AEI/10.13039/501100011033). The authors also thank the IAC High-Performance Computing support team for their contribution to the results of this research through the open-project ``burros'' initiative. This paper made use of the IAC HTCondor facility (\href{http://research.cs.wisc.edu/htcondor/}{http://research.cs.wisc.edu/htcondor/}), partly financed by the Ministry of Economy and Competitiveness with FEDER funds (IACA13-3E-2493). In addition to the codes explicitly cited in the text, this work made use of the following software packages: \textsc{matplotlib} \citep{2007CSE.....9...90H}, \textsc{numpy} \citep{2020Natur.585..357H}, and \textsc{scipy} \citep{2020NaMet..17..261V}. Finally, the authors thank the anonymous referee for the constructive feedback, which significantly improved the quality and clarity of this manuscript.
        }
    \end{acknowledgements}

        {
        \bibliographystyle{aa}   
        \bibliography{export-bibtex}

@ARTICLE{2026arXiv260402111M,
       author = {{Ma{\'\i}z Apell{\'a}niz}, J. and {Gamen}, R.~C. and {Holgado}, G. and {Rosu}, S. and {Arias}, J.~I. and {Sim{\'o}n-D{\'\i}az}, S. and {Pellerin}, A. and {Abdul-Masih}, M. and {Madero Fuentes}, E. and {Molina-Calzada}, J.~A. and {Barb{\'a}}, R.~H.},
        title = "{Multiplicity of Massive stars in the Milky Way (M3W). I. Project description, UNWIND, application to GLS 11 448, and DIB catalog}",
      journal = {arXiv e-prints},
         year = 2026,
        month = apr,
          eid = {arXiv:2604.02111},
        pages = {arXiv:2604.02111},
          doi = {10.48550/arXiv.2604.02111},
archivePrefix = {arXiv},
       eprint = {2604.02111},
 primaryClass = {astro-ph.SR},
       adsurl = {https://ui.adsabs.harvard.edu/abs/2026arXiv260402111M}
}

@ARTICLE{2025AJ....169..178H,
       author = {{Hubeny}, Ivan and {Bohlin}, Ralph and {Gordon}, Karl and {Fitzpatrick}, Edward},
        title = "{New Libraries of Nonlocal Thermodynamic Equilibrium Model Spectra for O and Early-B Stars}",
      journal = {\aj},
         year = 2025,
        month = mar,
       volume = {169},
       number = {3},
          eid = {178},
        pages = {178},
          doi = {10.3847/1538-3881/adb1bf},
       adsurl = {https://ui.adsabs.harvard.edu/abs/2025AJ....169..178H}
}

@ARTICLE{2024arXiv241016114S,
       author = {{Shenar}, Tomer},
        title = "{Observational constraints on massive binaries}",
      journal = {arXiv e-prints},
         year = 2024,
        month = oct,
          eid = {arXiv:2410.16114},
        pages = {arXiv:2410.16114},
          doi = {10.48550/arXiv.2410.16114},
archivePrefix = {arXiv},
       eprint = {2410.16114},
 primaryClass = {astro-ph.SR},
       adsurl = {https://ui.adsabs.harvard.edu/abs/2024arXiv241016114S}
}

@ARTICLE{2024arXiv240903329P,
       author = {{Puls}, Joachim and {Herrero}, Artemio and {Allende Prieto}, Carlos},
        title = "{Stellar Atmospheres}",
      journal = {arXiv e-prints},
         year = 2024,
        month = sep,
          eid = {arXiv:2409.03329},
        pages = {arXiv:2409.03329},
          doi = {10.48550/arXiv.2409.03329},
archivePrefix = {arXiv},
       eprint = {2409.03329},
 primaryClass = {astro-ph.SR},
       adsurl = {https://ui.adsabs.harvard.edu/abs/2024arXiv240903329P}
}

@ARTICLE{2024A&A...688A..97M,
       author = {{Montesinos}, Benjam{\'\i}n},
        title = "{Surface parameterisation and spectral synthesis of rapidly rotating stars. Vega as a testbed}",
      journal = {\aap},
         year = 2024,
        month = aug,
       volume = {688},
          eid = {A97},
        pages = {A97},
          doi = {10.1051/0004-6361/202449895},
archivePrefix = {arXiv},
       eprint = {2406.18392},
 primaryClass = {astro-ph.SR},
       adsurl = {https://ui.adsabs.harvard.edu/abs/2024A&A...688A..97M}
}

@ARTICLE{2024A&A...687A.228D,
       author = {{de Burgos}, A. and {Sim{\'o}n-D{\'\i}az}, S. and {Urbaneja}, M.~A. and {Puls}, J.},
        title = "{The IACOB project. X. Large-scale quantitative spectroscopic analysis of Galactic luminous blue stars}",
      journal = {\aap},
         year = 2024,
        month = jul,
       volume = {687},
          eid = {A228},
        pages = {A228},
          doi = {10.1051/0004-6361/202348808},
archivePrefix = {arXiv},
       eprint = {2312.00241},
 primaryClass = {astro-ph.SR},
       adsurl = {https://ui.adsabs.harvard.edu/abs/2024A&A...687A.228D}
}

@software{2024zndo..11397217H,
       author = {{HTCondor Team}},
        title = "{HTCondor}",
         year = 2024,
        month = may,
          eid = {10.5281/zenodo.11397217},
          doi = {10.5281/zenodo.11397217},
      version = {23.7.2},
    publisher = {Zenodo},
       adsurl = {https://ui.adsabs.harvard.edu/abs/2024zndo..11397217H}
}

@ARTICLE{2024A&A...685A..57L,
       author = {{Levenhagen}, Ronaldo S. and {Cur{\'e}}, Michel and {Arcos}, Catalina and {Diaz}, Marcos P. and {Araya}, Ignacio and {Am{\^o}res}, Eduardo B. and {Turis-Gallo}, Daniela and {Concha}, David},
        title = "{ZPEKTR: A code for spectral synthesis of fast-rotating stars}",
      journal = {\aap},
         year = 2024,
        month = may,
       volume = {685},
          eid = {A57},
        pages = {A57},
          doi = {10.1051/0004-6361/202348570},
       adsurl = {https://ui.adsabs.harvard.edu/abs/2024A&A...685A..57L}
}

@ARTICLE{2024MNRAS.530.1935S,
       author = {{Seeburger}, Rhys and {Rix}, Hans-Walter and {El-Badry}, Kareem and {Xiang}, Maosheng and {Fouesneau}, Morgan},
        title = "{Autonomous disentangling for spectroscopic surveys}",
      journal = {\mnras},
         year = 2024,
        month = may,
       volume = {530},
       number = {2},
        pages = {1935-1955},
          doi = {10.1093/mnras/stae982},
archivePrefix = {arXiv},
       eprint = {2405.19391},
 primaryClass = {astro-ph.SR},
       adsurl = {https://ui.adsabs.harvard.edu/abs/2024MNRAS.530.1935S}
}

@ARTICLE{2023A&A...669L..11A,
       author = {{Abdul-Masih}, M.},
        title = "{Effects of rotation on the spectroscopic observables of massive stars}",
      journal = {\aap},
         year = 2023,
        month = jan,
       volume = {669},
          eid = {L11},
        pages = {L11},
          doi = {10.1051/0004-6361/202245653},
archivePrefix = {arXiv},
       eprint = {2212.10485},
 primaryClass = {astro-ph.SR},
       adsurl = {https://ui.adsabs.harvard.edu/abs/2023A&A...669L..11A}
}

@ARTICLE{2021A&A...651A..96A,
       author = {{Abdul-Masih}, Michael and {Sana}, Hugues and {Hawcroft}, Calum and {Almeida}, Leonardo A. and {Brands}, Sarah A. and {de Mink}, Selma E. and {Justham}, Stephen and {Langer}, Norbert and {Mahy}, Laurent and {Marchant}, Pablo and {Menon}, Athira and {Puls}, Joachim and {Sundqvist}, Jon},
        title = "{Constraining the overcontact phase in massive binary evolution. I. Mixing in V382 Cyg, VFTS 352, and OGLE SMC-SC10 108086}",
      journal = {\aap},
         year = 2021,
        month = jul,
       volume = {651},
          eid = {A96},
        pages = {A96},
          doi = {10.1051/0004-6361/202040195},
archivePrefix = {arXiv},
       eprint = {2104.07621},
 primaryClass = {astro-ph.SR},
       adsurl = {https://ui.adsabs.harvard.edu/abs/2021A&A...651A..96A}
}

@ARTICLE{2021arXiv210402829H,
       author = {{Hubeny}, Ivan and {Allende Prieto}, Carlos and {Osorio}, Yeisson and {Lanz}, Thierry},
        title = "{TLUSTY and SYNSPEC Users's Guide IV: Upgraded Versions 208 and 54}",
      journal = {arXiv e-prints},
         year = 2021,
        month = apr,
          eid = {arXiv:2104.02829},
        pages = {arXiv:2104.02829},
          doi = {10.48550/arXiv.2104.02829},
archivePrefix = {arXiv},
       eprint = {2104.02829},
 primaryClass = {astro-ph.SR},
       adsurl = {https://ui.adsabs.harvard.edu/abs/2021arXiv210402829H}
}

@ARTICLE{2020A&A...642A.172P,
       author = {{Puls}, J. and {Najarro}, F. and {Sundqvist}, J.~O. and {Sen}, K.},
        title = "{Atmospheric NLTE models for the spectroscopic analysis of blue stars with winds. V. Complete comoving frame transfer, and updated modeling of X-ray emission}",
      journal = {\aap},
         year = 2020,
        month = oct,
       volume = {642},
          eid = {A172},
        pages = {A172},
          doi = {10.1051/0004-6361/202038464},
archivePrefix = {arXiv},
       eprint = {2011.02310},
 primaryClass = {astro-ph.SR},
       adsurl = {https://ui.adsabs.harvard.edu/abs/2020A&A...642A.172P}
}

@ARTICLE{2020Natur.585..357H,
       author = {{Harris}, Charles R. and {Millman}, K. Jarrod and {van der Walt}, St{\'e}fan J. and {Gommers}, Ralf and {Virtanen}, Pauli and {Cournapeau}, David and {Wieser}, Eric and {Taylor}, Julian and {Berg}, Sebastian and {Smith}, Nathaniel J. and {Kern}, Robert and {Picus}, Matti and {Hoyer}, Stephan and {van Kerkwijk}, Marten H. and {Brett}, Matthew and {Haldane}, Allan and {del R{\'\i}o}, Jaime Fern{\'a}ndez and {Wiebe}, Mark and {Peterson}, Pearu and {G{\'e}rard-Marchant}, Pierre and {Sheppard}, Kevin and {Reddy}, Tyler and {Weckesser}, Warren and {Abbasi}, Hameer and {Gohlke}, Christoph and {Oliphant}, Travis E.},
        title = "{Array programming with NumPy}",
      journal = {\nat},
         year = 2020,
        month = sep,
       volume = {585},
       number = {7825},
        pages = {357-362},
          doi = {10.1038/s41586-020-2649-2},
archivePrefix = {arXiv},
       eprint = {2006.10256},
 primaryClass = {cs.MS},
       adsurl = {https://ui.adsabs.harvard.edu/abs/2020Natur.585..357H}
}

@ARTICLE{2020A&A...636A..59A,
       author = {{Abdul-Masih}, Michael and {Sana}, Hugues and {Conroy}, Kyle E. and {Sundqvist}, Jon and {Pr{\v{s}}a}, Andrej and {Kochoska}, Angela and {Puls}, Joachim},
        title = "{Spectroscopic patch model for massive stars using PHOEBE II and FASTWIND}",
      journal = {\aap},
         year = 2020,
        month = apr,
       volume = {636},
          eid = {A59},
        pages = {A59},
          doi = {10.1051/0004-6361/201937341},
archivePrefix = {arXiv},
       eprint = {2003.09008},
 primaryClass = {astro-ph.SR},
       adsurl = {https://ui.adsabs.harvard.edu/abs/2020A&A...636A..59A}
}

@ARTICLE{2020NaMet..17..261V,
       author = {{Virtanen}, Pauli and {Gommers}, Ralf and {Oliphant}, Travis E. and {Haberland}, Matt and {Reddy}, Tyler and {Cournapeau}, David and {Burovski}, Evgeni and {Peterson}, Pearu and {Weckesser}, Warren and {Bright}, Jonathan and {van der Walt}, St{\'e}fan J. and {Brett}, Matthew and {Wilson}, Joshua and {Millman}, K. Jarrod and {Mayorov}, Nikolay and {Nelson}, Andrew R.~J. and {Jones}, Eric and {Kern}, Robert and {Larson}, Eric and {Carey}, C.~J. and {Polat}, {\.I}lhan and {Feng}, Yu and {Moore}, Eric W. and {VanderPlas}, Jake and {Laxalde}, Denis and {Perktold}, Josef and {Cimrman}, Robert and {Henriksen}, Ian and {Quintero}, E.~A. and {Harris}, Charles R. and {Archibald}, Anne M. and {Ribeiro}, Ant{\^o}nio H. and {Pedregosa}, Fabian and {van Mulbregt}, Paul and {SciPy 1.  0 Contributors}},
        title = "{SciPy 1.0: fundamental algorithms for scientific computing in Python}",
      journal = {Nature Methods},
         year = 2020,
        month = feb,
       volume = {17},
        pages = {261-272},
          doi = {10.1038/s41592-019-0686-2},
archivePrefix = {arXiv},
       eprint = {1907.10121},
 primaryClass = {cs.MS},
       adsurl = {https://ui.adsabs.harvard.edu/abs/2020NaMet..17..261V}
}

@ARTICLE{2019ApJ...880..115A,
       author = {{Abdul-Masih}, Michael and {Sana}, Hugues and {Sundqvist}, Jon and {Mahy}, Laurent and {Menon}, Athira and {Almeida}, Leonardo A. and {De Koter}, Alex and {de Mink}, Selma E. and {Justham}, Stephen and {Langer}, Norbert and {Puls}, Joachim and {Shenar}, Tomer and {Tramper}, Frank},
        title = "{Clues on the Origin and Evolution of Massive Contact Binaries: Atmosphere Analysis of VFTS 352}",
      journal = {\apj},
         year = 2019,
        month = aug,
       volume = {880},
       number = {2},
          eid = {115},
        pages = {115},
          doi = {10.3847/1538-4357/ab24d4},
archivePrefix = {arXiv},
       eprint = {1906.01066},
 primaryClass = {astro-ph.SR},
       adsurl = {https://ui.adsabs.harvard.edu/abs/2019ApJ...880..115A}
}

@ARTICLE{2019A&A...621A..85H,
       author = {{Hainich}, R. and {Ramachandran}, V. and {Shenar}, T. and {Sander}, A.~A.~C. and {Todt}, H. and {Gruner}, D. and {Oskinova}, L.~M. and {Hamann}, W.-R.},
        title = "{PoWR grids of non-LTE model atmospheres for OB-type stars of various metallicities}",
      journal = {\aap},
         year = 2019,
        month = jan,
       volume = {621},
          eid = {A85},
        pages = {A85},
          doi = {10.1051/0004-6361/201833787},
archivePrefix = {arXiv},
       eprint = {1811.06307},
 primaryClass = {astro-ph.SR},
       adsurl = {https://ui.adsabs.harvard.edu/abs/2019A&A...621A..85H}
}

@BOOK{2018maeb.book.....P,
       author = {{Pr{\v{s}}a}, Andrej},
        title = "{Modeling and Analysis of Eclipsing Binary Stars; The theory and design principles of PHOEBE}",
         year = 2018,
          doi = {10.1088/978-0-7503-1287-5},
       adsurl = {https://ui.adsabs.harvard.edu/abs/2018maeb.book.....P}
}

@ARTICLE{2018A&A...618A..25A,
       author = {{Allende Prieto}, C. and {Koesterke}, L. and {Hubeny}, I. and {Bautista}, M.~A. and {Barklem}, P.~S. and {Nahar}, S.~N.},
        title = "{A collection of model stellar spectra for spectral types B to early-M}",
      journal = {\aap},
         year = 2018,
        month = oct,
       volume = {618},
          eid = {A25},
        pages = {A25},
          doi = {10.1051/0004-6361/201732484},
archivePrefix = {arXiv},
       eprint = {1807.06049},
 primaryClass = {astro-ph.SR},
       adsurl = {https://ui.adsabs.harvard.edu/abs/2018A&A...618A..25A}
}

@ARTICLE{2017arXiv170601859H,
       author = {{Hubeny}, Ivan and {Lanz}, Thierry},
        title = "{A brief introductory guide to TLUSTY and SYNSPEC}",
      journal = {arXiv e-prints},
         year = 2017,
        month = jun,
          eid = {arXiv:1706.01859},
        pages = {arXiv:1706.01859},
          doi = {10.48550/arXiv.1706.01859},
archivePrefix = {arXiv},
       eprint = {1706.01859},
 primaryClass = {astro-ph.SR},
       adsurl = {https://ui.adsabs.harvard.edu/abs/2017arXiv170601859H}
}

@ARTICLE{2017arXiv170601935H,
       author = {{Hubeny}, Ivan and {Lanz}, Thierry},
        title = "{TLUSTY User's Guide II: Reference Manual}",
      journal = {arXiv e-prints},
         year = 2017,
        month = jun,
          eid = {arXiv:1706.01935},
        pages = {arXiv:1706.01935},
          doi = {10.48550/arXiv.1706.01935},
archivePrefix = {arXiv},
       eprint = {1706.01935},
 primaryClass = {astro-ph.SR},
       adsurl = {https://ui.adsabs.harvard.edu/abs/2017arXiv170601935H}
}

@ARTICLE{2017arXiv170601937H,
       author = {{Hubeny}, Ivan and {Lanz}, Thierry},
        title = "{TLUSTY User's Guide III: Operational Manual}",
      journal = {arXiv e-prints},
         year = 2017,
        month = jun,
          eid = {arXiv:1706.01937},
        pages = {arXiv:1706.01937},
          doi = {10.48550/arXiv.1706.01937},
archivePrefix = {arXiv},
       eprint = {1706.01937},
 primaryClass = {astro-ph.SR},
       adsurl = {https://ui.adsabs.harvard.edu/abs/2017arXiv170601937H}
}

@ARTICLE{2016ApJS..227...29P,
       author = {{Pr{\v{s}}a}, A. and {Conroy}, K.~E. and {Horvat}, M. and {Pablo}, H. and {Kochoska}, A. and {Bloemen}, S. and {Giammarco}, J. and {Hambleton}, K.~M. and {Degroote}, P.},
        title = "{Physics Of Eclipsing Binaries. II. Toward the Increased Model Fidelity}",
      journal = {\apjs},
         year = 2016,
        month = dec,
       volume = {227},
       number = {2},
          eid = {29},
        pages = {29},
          doi = {10.3847/1538-4365/227/2/29},
archivePrefix = {arXiv},
       eprint = {1609.08135},
 primaryClass = {astro-ph.SR},
       adsurl = {https://ui.adsabs.harvard.edu/abs/2016ApJS..227...29P}
}

@INPROCEEDINGS{2016AAS...22734418G,
       author = {{Gropp}, Jeffrey D. and {Prsa}, Andrej},
        title = "{The Reflection Effect in Eclipsing Binaries}",
    booktitle = {American Astronomical Society Meeting Abstracts \#227},
         year = 2016,
       series = {American Astronomical Society Meeting Abstracts},
       volume = {227},
        month = jan,
          eid = {344.18},
        pages = {344.18},
       adsurl = {https://ui.adsabs.harvard.edu/abs/2016AAS...22734418G}
}

@BOOK{2015tsaa.book.....H,
       author = {{Hubeny}, Ivan and {Mihalas}, Dimitri},
        title = "{Theory of Stellar Atmospheres. An Introduction to Astrophysical Non-equilibrium Quantitative Spectroscopic Analysis}",
         year = 2015,
       adsurl = {https://ui.adsabs.harvard.edu/abs/2015tsaa.book.....H}
}

@ARTICLE{2013A&A...552A..39P,
       author = {{Palate}, M. and {Rauw}, G. and {Koenigsberger}, G. and {Moreno}, E.},
        title = "{Spectral modelling of massive binary systems}",
      journal = {\aap},
         year = 2013,
        month = apr,
       volume = {552},
          eid = {A39},
        pages = {A39},
          doi = {10.1051/0004-6361/201219754},
archivePrefix = {arXiv},
       eprint = {1302.5201},
 primaryClass = {astro-ph.SR},
       adsurl = {https://ui.adsabs.harvard.edu/abs/2013A&A...552A..39P}
}

@ARTICLE{2012AJ....144..120M,
       author = {{M{\'e}sz{\'a}ros}, Sz. and {Allende Prieto}, C. and {Edvardsson}, B. and {Castelli}, F. and {Garc{\'\i}a P{\'e}rez}, A.~E. and {Gustafsson}, B. and {Majewski}, S.~R. and {Plez}, B. and {Schiavon}, R. and {Shetrone}, M. and {de Vicente}, A.},
        title = "{New ATLAS9 and MARCS Model Atmosphere Grids for the Apache Point Observatory Galactic Evolution Experiment (APOGEE)}",
      journal = {\aj},
         year = 2012,
        month = oct,
       volume = {144},
       number = {4},
          eid = {120},
        pages = {120},
          doi = {10.1088/0004-6256/144/4/120},
archivePrefix = {arXiv},
       eprint = {1208.1916},
 primaryClass = {astro-ph.SR},
       adsurl = {https://ui.adsabs.harvard.edu/abs/2012AJ....144..120M}
}

@ARTICLE{2012ARA&A..50..107L,
       author = {{Langer}, N.},
        title = "{Presupernova Evolution of Massive Single and Binary Stars}",
      journal = {\araa},
         year = 2012,
        month = sep,
       volume = {50},
        pages = {107-164},
          doi = {10.1146/annurev-astro-081811-125534},
archivePrefix = {arXiv},
       eprint = {1206.5443},
 primaryClass = {astro-ph.SR},
       adsurl = {https://ui.adsabs.harvard.edu/abs/2012ARA&A..50..107L}
}

@ARTICLE{2012MNRAS.425...21T,
       author = {{Tennyson}, Jonathan and {Yurchenko}, Sergei N.},
        title = "{ExoMol: molecular line lists for exoplanet and other atmospheres}",
      journal = {\mnras},
         year = 2012,
        month = sep,
       volume = {425},
       number = {1},
        pages = {21-33},
          doi = {10.1111/j.1365-2966.2012.21440.x},
archivePrefix = {arXiv},
       eprint = {1204.0124},
 primaryClass = {astro-ph.EP},
       adsurl = {https://ui.adsabs.harvard.edu/abs/2012MNRAS.425...21T}
}

@ARTICLE{2012A&A...537A.146E,
       author = {{Ekstr{\"o}m}, S. and {Georgy}, C. and {Eggenberger}, P. and {Meynet}, G. and {Mowlavi}, N. and {Wyttenbach}, A. and {Granada}, A. and {Decressin}, T. and {Hirschi}, R. and {Frischknecht}, U. and {Charbonnel}, C. and {Maeder}, A.},
        title = "{Grids of stellar models with rotation. I. Models from 0.8 to 120 M$_{☉}$ at solar metallicity (Z = 0.014)}",
      journal = {\aap},
         year = 2012,
        month = jan,
       volume = {537},
          eid = {A146},
        pages = {A146},
          doi = {10.1051/0004-6361/201117751},
archivePrefix = {arXiv},
       eprint = {1110.5049},
 primaryClass = {astro-ph.SR},
       adsurl = {https://ui.adsabs.harvard.edu/abs/2012A&A...537A.146E}
}

@ARTICLE{2011A&A...536A..58R,
       author = {{Rivero Gonz{\'a}lez}, J.~G. and {Puls}, J. and {Najarro}, F.},
        title = "{Nitrogen line spectroscopy of O-stars. I. Nitrogen III emission line formation revisited}",
      journal = {\aap},
         year = 2011,
        month = dec,
       volume = {536},
          eid = {A58},
        pages = {A58},
          doi = {10.1051/0004-6361/201117101},
archivePrefix = {arXiv},
       eprint = {1109.3595},
 primaryClass = {astro-ph.SR},
       adsurl = {https://ui.adsabs.harvard.edu/abs/2011A&A...536A..58R}
}

@ARTICLE{2011A&A...533A..43E,
       author = {{Espinosa Lara}, F. and {Rieutord}, M.},
        title = "{Gravity darkening in rotating stars}",
      journal = {\aap},
         year = 2011,
        month = sep,
       volume = {533},
          eid = {A43},
        pages = {A43},
          doi = {10.1051/0004-6361/201117252},
archivePrefix = {arXiv},
       eprint = {1109.3038},
 primaryClass = {astro-ph.SR},
       adsurl = {https://ui.adsabs.harvard.edu/abs/2011A&A...533A..43E}
}

@ARTICLE{2010AJ....139.1628D,
       author = {{Doi}, Mamoru and {Tanaka}, Masayuki and {Fukugita}, Masataka and {Gunn}, James E. and {Yasuda}, Naoki and {Ivezi{\'c}}, {\v{Z}}eljko and {Brinkmann}, Jon and {de Haars}, Ernst and {Kleinman}, S.~J. and {Krzesinski}, Jurek and {French Leger}, R.},
        title = "{Photometric Response Functions of the Sloan Digital Sky Survey Imager}",
      journal = {\aj},
         year = 2010,
        month = apr,
       volume = {139},
       number = {4},
        pages = {1628-1648},
          doi = {10.1088/0004-6256/139/4/1628},
archivePrefix = {arXiv},
       eprint = {1002.3701},
 primaryClass = {astro-ph.IM},
       adsurl = {https://ui.adsabs.harvard.edu/abs/2010AJ....139.1628D}
}

@ARTICLE{2009ARA&A..47..481A,
       author = {{Asplund}, Martin and {Grevesse}, Nicolas and {Sauval}, A. Jacques and {Scott}, Pat},
        title = "{The Chemical Composition of the Sun}",
      journal = {\araa},
         year = 2009,
        month = sep,
       volume = {47},
       number = {1},
        pages = {481-522},
          doi = {10.1146/annurev.astro.46.060407.145222},
archivePrefix = {arXiv},
       eprint = {0909.0948},
 primaryClass = {astro-ph.SR},
       adsurl = {https://ui.adsabs.harvard.edu/abs/2009ARA&A..47..481A}
}

@ARTICLE{2009A&A...494..399H,
       author = {{Hadrava}, P.},
        title = "{Notes on the disentangling of spectra. I. Enhancement in precision}",
      journal = {\aap},
         year = 2009,
        month = jan,
       volume = {494},
       number = {1},
        pages = {399-402},
          doi = {10.1051/0004-6361:200810810},
archivePrefix = {arXiv},
       eprint = {0901.4262},
 primaryClass = {astro-ph.IM},
       adsurl = {https://ui.adsabs.harvard.edu/abs/2009A&A...494..399H}
}

@BOOK{2009pfer.book.....M,
       author = {{Maeder}, Andr{\'e}},
        title = "{Physics, Formation and Evolution of Rotating Stars}",
         year = 2009,
          doi = {10.1007/978-3-540-76949-1},
       adsurl = {https://ui.adsabs.harvard.edu/abs/2009pfer.book.....M}
}

@BOOK{2008oasp.book.....G,
       author = {{Gray}, David F.},
        title = "{The Observation and Analysis of Stellar Photospheres}",
         year = 2008,
       adsurl = {https://ui.adsabs.harvard.edu/abs/2008oasp.book.....G}
}

@ARTICLE{2007ApJS..169...83L,
       author = {{Lanz}, Thierry and {Hubeny}, Ivan},
        title = "{A Grid of NLTE Line-blanketed Model Atmospheres of Early B-Type Stars}",
      journal = {\apjs},
         year = 2007,
        month = mar,
       volume = {169},
       number = {1},
        pages = {83-104},
          doi = {10.1086/511270},
archivePrefix = {arXiv},
       eprint = {astro-ph/0611891},
 primaryClass = {astro-ph},
       adsurl = {https://ui.adsabs.harvard.edu/abs/2007ApJS..169...83L}
}

@ARTICLE{2007CSE.....9...90H,
       author = {{Hunter}, John D.},
        title = "{Matplotlib: A 2D Graphics Environment}",
      journal = {Computing in Science and Engineering},
         year = 2007,
        month = jan,
       volume = {9},
       number = {3},
        pages = {90-95},
          doi = {10.1109/MCSE.2007.55},
       adsurl = {https://ui.adsabs.harvard.edu/abs/2007CSE.....9...90H}
}

@ARTICLE{2005A&A...435..669P,
       author = {{Puls}, J. and {Urbaneja}, M.~A. and {Venero}, R. and {Repolust}, T. and {Springmann}, U. and {Jokuthy}, A. and {Mokiem}, M.~R.},
        title = "{Atmospheric NLTE-models for the spectroscopic analysis of blue stars with winds. II. Line-blanketed models}",
      journal = {\aap},
         year = 2005,
        month = may,
       volume = {435},
       number = {2},
        pages = {669-698},
          doi = {10.1051/0004-6361:20042365},
archivePrefix = {arXiv},
       eprint = {astro-ph/0411398},
 primaryClass = {astro-ph},
       adsurl = {https://ui.adsabs.harvard.edu/abs/2005A&A...435..669P}
}

@ARTICLE{2005MSAIS...8...34C,
       author = {{Castelli}, F.},
        title = "{DFSYNTHE: how to use it}",
      journal = {Memorie della Societa Astronomica Italiana Supplementi},
         year = 2005,
        month = jan,
       volume = {8},
        pages = {34},
       adsurl = {https://ui.adsabs.harvard.edu/abs/2005MSAIS...8...34C}
}

@ARTICLE{2005MSAIS...8...14K,
       author = {{Kurucz}, Robert L.},
        title = "{ATLAS12, SYNTHE, ATLAS9, WIDTH9, et cetera}",
      journal = {Memorie della Societa Astronomica Italiana Supplementi},
         year = 2005,
        month = jan,
       volume = {8},
        pages = {14},
       adsurl = {https://ui.adsabs.harvard.edu/abs/2005MSAIS...8...14K}
}

@ARTICLE{2003ApJS..146..417L,
       author = {{Lanz}, Thierry and {Hubeny}, Ivan},
        title = "{A Grid of Non-LTE Line-blanketed Model Atmospheres of O-Type Stars}",
      journal = {\apjs},
         year = 2003,
        month = jun,
       volume = {146},
       number = {2},
        pages = {417-441},
          doi = {10.1086/374373},
archivePrefix = {arXiv},
       eprint = {astro-ph/0210157},
 primaryClass = {astro-ph},
       adsurl = {https://ui.adsabs.harvard.edu/abs/2003ApJS..146..417L}
}

@INPROCEEDINGS{2003IAUS..210P.A20C,
       author = {{Castelli}, F. and {Kurucz}, R.~L.},
        title = "{New Grids of ATLAS9 Model Atmospheres}",
    booktitle = {Modelling of Stellar Atmospheres},
         year = 2003,
       editor = {{Piskunov}, N. and {Weiss}, W.~W. and {Gray}, D.~F.},
       series = {IAU Symposium},
       volume = {210},
        month = jan,
        pages = {A20},
          doi = {10.48550/arXiv.astro-ph/0405087},
archivePrefix = {arXiv},
       eprint = {astro-ph/0405087},
 primaryClass = {astro-ph},
       adsurl = {https://ui.adsabs.harvard.edu/abs/2003IAUS..210P.A20C}
}

@ARTICLE{2000ARA&A..38..613K,
       author = {{Kudritzki}, Rolf-Peter and {Puls}, Joachim},
        title = "{Winds from Hot Stars}",
      journal = {\araa},
         year = 2000,
        month = jan,
       volume = {38},
        pages = {613-666},
          doi = {10.1146/annurev.astro.38.1.613},
       adsurl = {https://ui.adsabs.harvard.edu/abs/2000ARA&A..38..613K}
}

@ARTICLE{1999A&A...347..185M,
       author = {{Maeder}, Andr{\'e}},
        title = "{Stellar evolution with rotation IV: von Zeipel's theorem and anisotropic losses of mass and angular momentum}",
      journal = {\aap},
         year = 1999,
        month = jul,
       volume = {347},
        pages = {185-193},
       adsurl = {https://ui.adsabs.harvard.edu/abs/1999A&A...347..185M}
}

@ARTICLE{1998SSRv...85..161G,
       author = {{Grevesse}, N. and {Sauval}, A.~J.},
        title = "{Standard Solar Composition}",
      journal = {\ssr},
         year = 1998,
        month = may,
       volume = {85},
        pages = {161-174},
          doi = {10.1023/A:1005161325181},
       adsurl = {https://ui.adsabs.harvard.edu/abs/1998SSRv...85..161G}
}

@ARTICLE{1997A&A...323..488S,
       author = {{Santolaya-Rey}, A.~E. and {Puls}, J. and {Herrero}, A.},
        title = "{Atmospheric NLTE-models for the spectroscopic analysis of luminous blue stars with winds.}",
      journal = {\aap},
         year = 1997,
        month = jul,
       volume = {323},
        pages = {488-512},
       adsurl = {https://ui.adsabs.harvard.edu/abs/1997A&A...323..488S}
}

@ARTICLE{1996AJ....111.1748F,
       author = {{Fukugita}, M. and {Ichikawa}, T. and {Gunn}, J.~E. and {Doi}, M. and {Shimasaku}, K. and {Schneider}, D.~P.},
        title = "{The Sloan Digital Sky Survey Photometric System}",
      journal = {\aj},
         year = 1996,
        month = apr,
       volume = {111},
        pages = {1748},
          doi = {10.1086/117915},
       adsurl = {https://ui.adsabs.harvard.edu/abs/1996AJ....111.1748F}
}

@ARTICLE{1995ApJ...439..875H,
       author = {{Hubeny}, I. and {Lanz}, T.},
        title = "{Non-LTE Line-blanketed Model Atmospheres of Hot Stars. I. Hybrid Complete Linearization/Accelerated Lambda Iteration Method}",
      journal = {\apj},
         year = 1995,
        month = feb,
       volume = {439},
        pages = {875},
          doi = {10.1086/175226},
       adsurl = {https://ui.adsabs.harvard.edu/abs/1995ApJ...439..875H}
}

@ARTICLE{1979ApJS...40....1K,
       author = {{Kurucz}, R.~L.},
        title = "{Model atmospheres for G, F, A, B, and O stars.}",
      journal = {\apjs},
         year = 1979,
        month = may,
       volume = {40},
        pages = {1-340},
          doi = {10.1086/190589},
       adsurl = {https://ui.adsabs.harvard.edu/abs/1979ApJS...40....1K}
}

@ARTICLE{1933MNRAS..93..478C,
       author = {{Carroll}, J.~A.},
        title = "{The spectroscopic determination of stellar rotation and its effect on line profiles}",
      journal = {\mnras},
         year = 1933,
        month = may,
       volume = {93},
        pages = {478-507},
          doi = {10.1093/mnras/93.7.478},
       adsurl = {https://ui.adsabs.harvard.edu/abs/1933MNRAS..93..478C}
}

@ARTICLE{1924MNRAS..84..665V,
       author = {{von Zeipel}, H.},
        title = "{The radiative equilibrium of a rotating system of gaseous masses}",
      journal = {\mnras},
         year = 1924,
        month = jun,
       volume = {84},
        pages = {665-683},
          doi = {10.1093/mnras/84.9.665},
       adsurl = {https://ui.adsabs.harvard.edu/abs/1924MNRAS..84..665V}
}
        }
    }

\begin{appendix}
    {
    \onecolumn
    \nolinenumbers
    
    \section{$\Teff-\logg$ plane for the adopted model atmosphere grids} \label{A: grids_parameters}
        {
        \begin{figure*}[ht!]
            \centering
            \resizebox{\hsize}{!}{\includegraphics[scale=1]{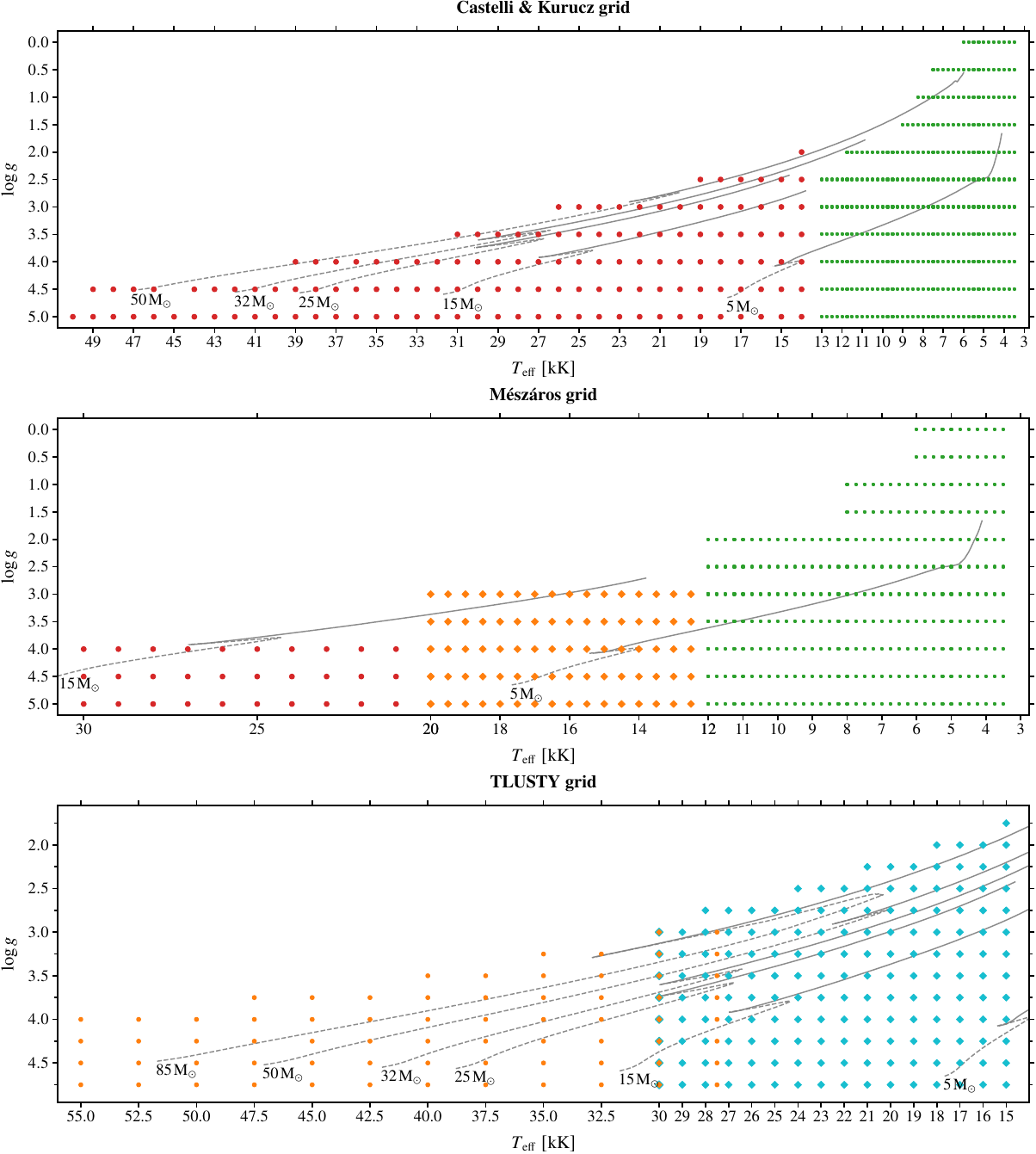}}
            \caption{
                Distribution of stellar model atmospheres in the $\Teff-\logg$ plane for the Kurucz and TLUSTY grids. Top panel: LTE-ATLAS9 grid from \citet{2003IAUS..210P.A20C}. The green squares indicate models in the low-temperature regime, with a sampling of $\Delta \Teff = 250\,\K$. The red circles represent high-temperature models with $\Delta \Teff = 1000\,\K$. Middle panel: Same as the top panel but with the LTE-ATLAS9 grid computed by \citet{2012AJ....144..120M}. The orange diamonds display models in the middle-temperature regime with $\Delta \Teff = 500\,\K$. Bottom panel: Non-LTE TLUSTY-based grid. The blue diamonds denote the BSTAR2006 models with a temperature sampling of $\Delta \Teff = 1000\,\K$ \citep{2007ApJS..169...83L}. The orange circles represent the OSTAR2002 grid with $\Delta \Teff = 2500\,\K$ \citep{2003ApJS..146..417L}. The gray lines show the Geneva stellar evolution tracks at solar metallicity: dashed during the core hydrogen-burning phase, and solid when the hydrogen-helium burning is inactive \citep{2012A&A...537A.146E}.
            }
            \label{Fig: grids_parameters}
        \end{figure*}
        }
    \FloatBarrier

    \newpage
    \section{Convergence of the emergent flux as a function of the $\mu$-grid resolution} \label{A: mu_convergence}
        {
        \begin{figure*}[th!]
            \centering
            \resizebox{\hsize}{!}{\includegraphics{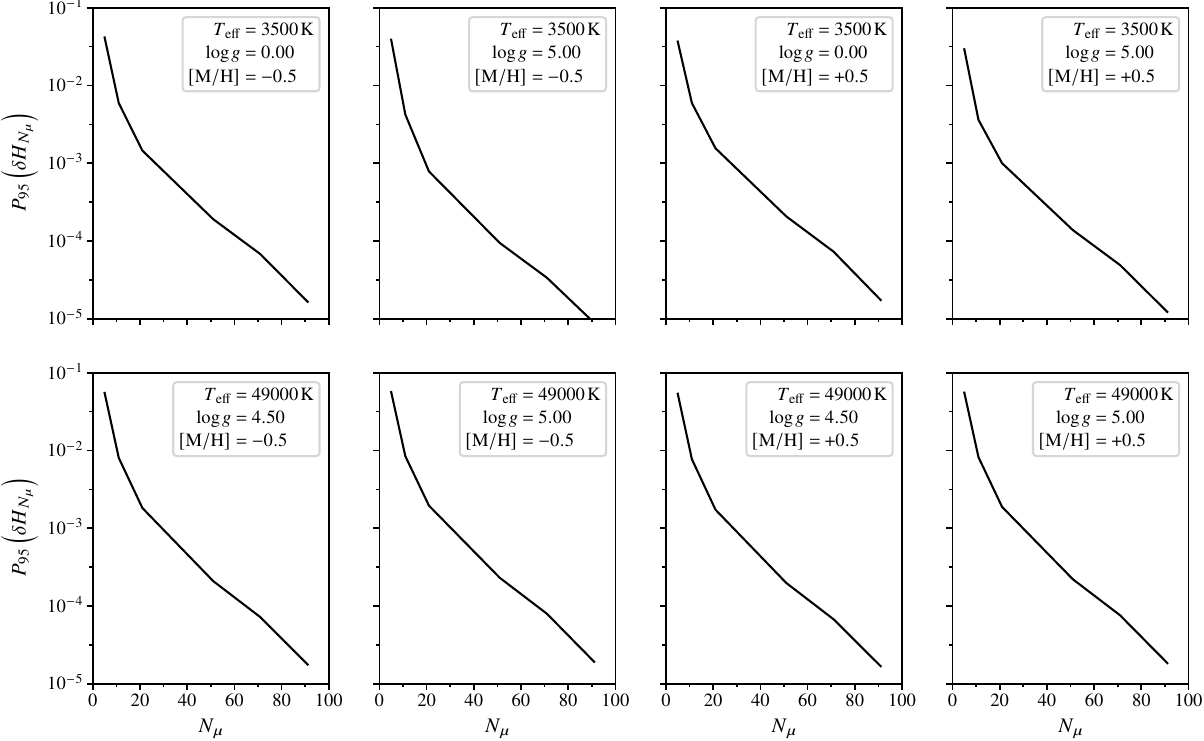}}
            \caption{
                Convergence of the emergent Eddington flux as a function of the total number of $\mu$ points, $N_{\mu  \equiv \cos{\theta}}$. The curves show the 95th percentile of the relative flux difference, $P_{95}\left(\delta H\right)$, computed over the $3000-9000\,\AA$ range and using $N_{\mu}=101$ as the reference solution (Eqs.~\ref{Eq: mu_analysis1}~and~\ref{Eq: mu_analysis2}). The vertical axis is shown on a logarithmic scale.
            }
            \label{Fig: mu_convergence}
        \end{figure*}

        \begin{multicols}{2}
            In this section, we quantify the impact of angular resolution on the numerical computation of stellar fluxes. To this end, we calculated specific intensities, $I\left(\lambda, \mu \equiv \cos{\theta}\right)$, with \SYNSPEC{} for a different amount of $\mu$ points: $N_{\mu}~=~5$,~$11$,~$21$,~$51$,~$71$,~$91$,~and~$101$. The calculations were performed for representative models of the Castelli\&Kurucz grid, spanning a wide range of effective temperatures, surface gravities, and metallicities in the $3000-9000\,\AA$ wavelength interval \citep{2003IAUS..210P.A20C}. For each configuration, we computed the corresponding Eddington flux, $H_{N_{\mu}}\left(\lambda\right)$, by numerical integration of the resulting specific intensities (Eq.~\ref{Eq: Edd_flux}).

            To evaluate convergence, we compared the different $H_{N_{\mu}}\left(\lambda\right)$ fluxes. The reference flux is defined using $N_{\mu}=101$, which corresponds to a sufficiently dense angular discretization such that the numerical $\mu$-integral is fully converged. Thus, the relative flux difference is given by
            \vspace{0.2cm}
            \begin{equation}
                \delta H_{N_{\mu}}\left(\lambda\right) =
                                    \left| \frac{H_{N_{\mu}}\left(\lambda\right)-H_{101}\left(\lambda\right)}{H_{101}\left(\lambda\right)} \right|,
                \label{Eq: mu_analysis1}
            \end{equation}
            \vspace{0.2cm}
            and is summarized through the 95th percentile of its wavelength distribution:
            \vspace{0.2cm}
            \begin{equation}
                P_{95}\left(\delta H_{N_{\mu}}\right) =
                                                        Q_{0.95}\left[\delta H_{N_{\mu}}\left(\lambda\right)\right].
                \label{Eq: mu_analysis2}
            \end{equation}

            The resulting convergence curves are shown in Fig.~\ref{Fig: mu_convergence}. For all stellar parameters considered, the flux difference decreases monotonically with increasing $N_{\mu}$ -- indicating stable and well-behaved convergence. When expressed on logarithmic scale, the convergence is approximately linear over $N_{\mu}$, consistent with an exponential decrease with angular resolution.
            
            The largest flux differences occur in the low-resolution regime. For $N_{\mu} = 5$, the angular integration is insufficient to accurately reproduce the limb-darkening effect, yielding $P_{95}\left(\delta H_{N_{\mu}}\right) \approx 5\%$. Increasing the angular resolution to $N_{\mu} = 11$ reduces the discrepancies to approximately $1\%$. At $N_{\mu} = 21$, $P_{95}\left(\delta H_{N_{\mu}}\right)$ falls below $0.1\%$ for all models. For $N_{\mu} = 51$, the solution approaches the convergence regime, with typical deviations below $0.01\%$. Further increasing the angular resolution to $N_{\mu}=71$ and $91$ yields only negligible improvements, with flux differences remaining between $0.01\%$ and $0.001\%$.

            Based on these results, we adopt $N_{\mu} = 101$ as the standard angular discretization for the intensity grids presented in this work. Although the convergence analysis shows that values of $N_\mu > 50$ already yield converged fluxes, adopting $N_{\mu}=101$ provides a conservative resolution for all model atmospheres. At this sampling, the numerical integration is fully converged and completely removes angular resolution as a potential source of systematic error.
        \end{multicols}
        }
    \FloatBarrier
    
    \newpage
    \section{Surface gravity distributions in a rapidly rotating B-type star} \label{A: rotatingBstar}
        {
        \begin{figure*}[th!]
            \centering
            \resizebox{\hsize}{!}{\includegraphics{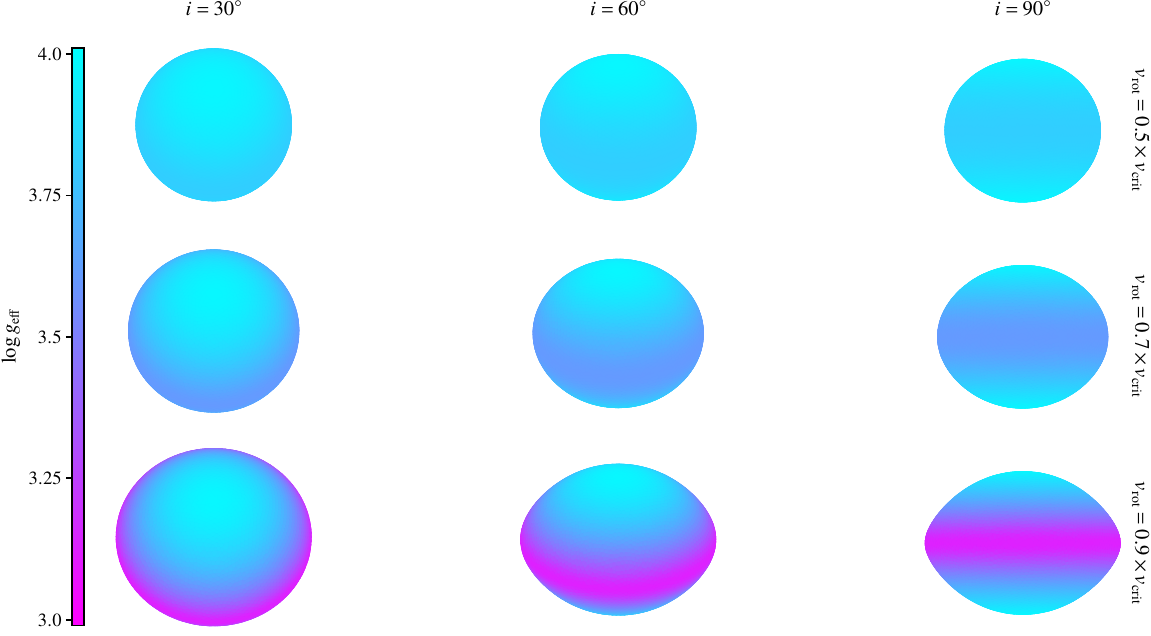}}
            \caption{
                Effective gravity distribution across the surface for a rotating B-type star with $\Teff = 21000\,\K$, $\Mstar = 7\,\Msol$, and $\Rpole = 4.5\,\Rsol$. The vertical color bar indicates the local surface gravity: darker and lighter colors represent lower and higher values, respectively. Rows: Models computed at $50\%$, $70\%$, and $90\%$ of the critical equatorial velocity, $\vrotcrit$. Columns: Different inclination angles of the rotation axis ($i = 30\deg$,~$60\deg$,~$90\deg$).
            }
            \label{Fig: rotatingBstar_loggeff}
        \end{figure*} 
        }
    }
\end{appendix}
\end{document}